\documentclass[reqno]{article}
\usepackage{ae} 
\usepackage[T1]{fontenc}
\usepackage[ansinew]{inputenc}
\usepackage{amsmath}
\usepackage{amssymb}
\usepackage{graphicx}
\usepackage{color}
\usepackage[colorlinks]{hyperref}
\usepackage{epsfig}
\usepackage{graphicx}

\newcommand{\beq}{\begin{equation}}
\newcommand{\eeq}{\end{equation}}
\newcommand\beqa{\begin{eqnarray}}
\newcommand\eeqa{\end{eqnarray}}
\newcommand\bea{\begin{array}}
\newcommand\eea{\end{array}}
\newcommand\ba{\begin{array}}
\newcommand\ea{\end{array}}
\newcommand{\nn}{\nonumber}

\newcommand{\neqa}{\nonumber\end{eqnarray}}
\newcommand{\la}[1]{\label{#1}}

\newcommand{\eq}[1]{(\ref{#1})}

\renewcommand{\d}{\partial}
\renewcommand{\O}{{\cal O}}

\newcommand{\<}{{\langle}}
\renewcommand{\>}{{\rangle}}

\newcommand{\re}{\relax{\rm I\kern-.18em R}}

\def\su2{{SU(2)}}

\def\tr{{\rm tr~}}

\def\one{{\bf 1}}

\def\schi{{\hat\chi}}

\def\sw{{\hat w}}

\def\O{{\Omega}}
\def\a{{\alpha}}
\def\({\left(}
\def\){\right)}
\def\[{\left[}
\def\]{\right]}

\def\l{\lambda}

\def\a{\alpha}
\def\b{\beta}

\def\th{\theta}

\def\<{\langle}
\def\>{\rangle}

\def\sdet{{\rm sdet}~}
\def\cP{{\cal P}}
\def\hD{\hat D}

\def\i2{\frac{i}{2}}

\begin{document}

\renewcommand{\thefootnote}{\fnsymbol{footnote}}
\setcounter{footnote}{0}

\thispagestyle{empty}
\begin{flushright}
LPTENS-07/53\\
\texttt{{\it Dedicated to the memory of Alexey Zamolodchikov}}\\
\end{flushright}
\vspace{1.0cm} \setcounter{footnote}{0}
\begin{center}
{\Large{\bf From Characters to Quantum (Super)Spin Chains via Fusion}\\
   }\vspace{8mm}
{\large  Vladimir~Kazakov$^{a,\!}$\footnote{Membre de l'Institut
Universitaire de France},\hspace{1cm} Pedro~Vieira\rm $^{a,b}$\\[7mm]
\large\it\small ${}^a$ Laboratoire de Physique Th\'eorique\\
de l'Ecole Normale Sup\'erieure et l'Universit\'e Paris-VI,\\
24 rue Lhomond, Paris  CEDEX 75231, France\footnote{ \tt\noindent
Email:\indent   kazakov@physique.ens.fr,
 \indent  pedrogvieira@gmail.com}
 \vspace{3mm}\\
\large\it\small ${}^b$  Departamento de F\'\i sica e Centro de F\'\i
sica do Porto\\ Faculdade de Ci\^encias da Universidade do Porto\\
Rua do Campo Alegre, 687, \,4169-007 Porto, Portugal; }

\end{center}
\noindent\\[10mm]
\begin{center}
{\sc Abstract}\\[2mm]
\end{center}

We give an elementary proof of the Bazhanov-Reshetikhin determinant formula
for  rational  transfer matrices of the twisted quantum super-spin chains
associated with the $gl(K|M)$ algebra. This formula describes the most
general  fusion of transfer matrices in symmetric representations into
arbitrary finite dimensional representations of the algebra and is at the
heart of analytical Bethe ansatz approach.
 Our technique represents a systematic generalization of the usual
Jacobi-Trudi formula for characters to its quantum analogue using certain
group derivatives.

\newpage
\tableofcontents
\newpage

\setcounter{page}{1}
\renewcommand{\thefootnote}{\arabic{footnote}}
\setcounter{footnote}{0}


\section{Introduction}

\begin{figure}
\centering \resizebox{150mm}{!}{\includegraphics{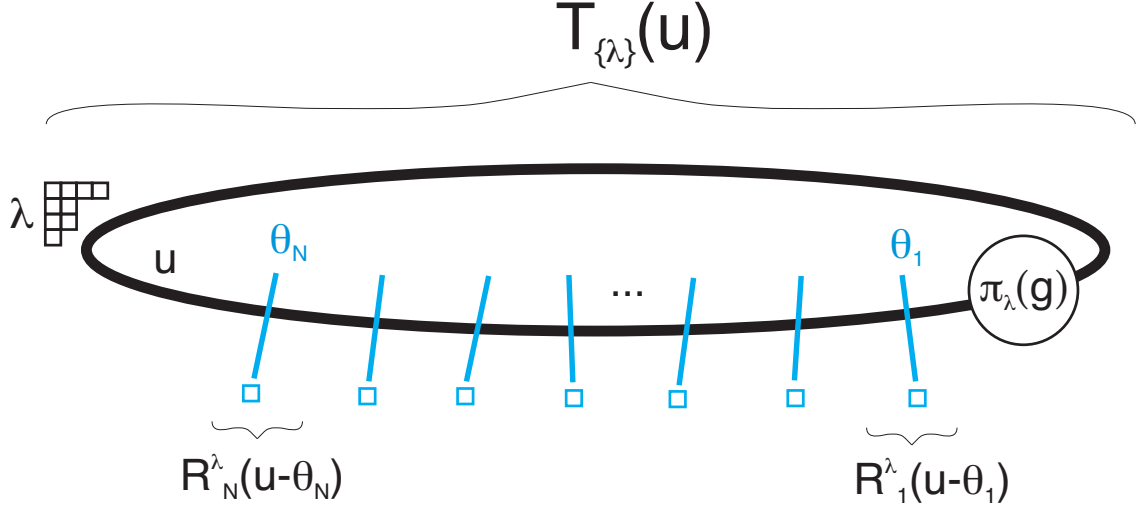}}
\caption{ \footnotesize    The central object of the paper,
transfer-matrix $T_{\{\l\}}(u)=\tr_\l
\(R_{N}^{\lambda}(u-\theta_N)\dots
R_{1}^{\lambda}(u-\theta_1)\pi_\l(g)\)$: the individual $R$-matrices
are multiplied along the auxiliary, horizontal space (solid circle)
of an arbitrary finite dimensional representation $\l$, represented
by its Young tableau, whereas the vertical lines represent the
spaces on which individual spins in fundamental representations act.
Each crossing corresponds to one $R$-matrix depending on a spectral
parameter $u-\theta_k$. The twist matrix $g$ is also taken in the
representation $\l$. The indices of the auxiliary space disappear
when taking the trace but this object is still a complex operator in
the quantum space with indices $\[T_{\{ \l \}}(u)\]^{i_1\dots i_N}_{j_1\dots
j_N}$.}\label{fig:Intro}
\end{figure}

Integrability of quantum spin chains was first realized in the famous
H.~Bethe solution \cite{Bethe} for the Heisenberg spin chain defined by the
Hamiltonian
\begin{equation}\label{HEIS}
    \hat H_{xxx}=\sum_{n=1}^N P_{n,n+1}
\end{equation}
where $P_{n,n+1}$ is the permutation operator. After a long development,
which has taken a few dozens of years, a much more general underlying
integrable structure was formulated in terms of the Yang-Baxter (YB), or
triangle relations for a very useful object: the  $R$-matrix. For the
$gl(K)$ algebra, the basic rational $R$-matrix is defined in the product
$V^{(K)}\otimes V^{(K)}$ of two $K$-dimensional vector spaces as follows
\begin{eqnarray*}
    R (u)&=&u\one\otimes\one +2P\\
    &=&u\one\otimes\one +2\sum_{\a,\b=1}^K e_{\b\a}  \otimes  e_{\a\b}
    \end{eqnarray*}
where $\[\,e_{\a\b}\]^{i}_{j}=\delta^i_\a\delta_{j,\b}$ are the generators
of $GL(K)$ in the fundamental representation and $P$ is the permutation
operator defined through the relation $P\(A\otimes B\)=\(B\otimes A\)P$.

There exists a more general $gl(K)$ $R$-matrix \cite{KulSk} satisfying the
YB relations, which acts in the tensor product of the fundamental
representation of the quantum spin, $V^{(K)}$, with a vector space on which
the representation $\l$ lives
\begin{equation}
    R^\lambda (u)=u+2\sum_{\a\b} e_{\b\a}  \otimes
    \pi_\l(e_{\a\b}), \la{RMATR}
\end{equation}
where $ \pi_\l(e_{\a\b})$ are the same generators in the representation
$\lambda$.  The general representation $\l$  of $gl(K)$ is defined by its
highest weight components $\l=(\l_1,\dots,\l_{K-1})$.

The next useful  object is the twisted transfer matrix
\begin{equation}\label{TMATR}
    T_{\{\lambda\}}(u)={\rm tr}_\lambda  \(R_{N}^\lambda(u-\theta_L)\otimes \dots
    \otimes R_{1}^\lambda(u-\theta_1)\, \pi_\lambda(g) \)
\end{equation}
where $g\in GL(K)$ is called the twist matrix. The trace goes over the
auxiliary space $\lambda$ and each $R$-matrix acts on the tensor product of
this space with the vector space associated with one of the sites of the
spin chain, indicated by the subscript, as depicted in figure
\ref{fig:Intro}. The rapidities $\theta_j$ are arbitrary constants. They appear naturally as the rapidities of physical particles when we interpret the $R$-matrices as scattering matrices. In the spin chain language they correspond to the generalization of the usual homogeneous spin chains to the inhomogeneous case.

Using the YB relations, one can check
\cite{KulResh1,Cherednik} that the transfer matrices commute for different
spectral parameters and different representations\footnote{In section
\ref{Rcomsec} we review the fusion procedure and prove that for symmetric
representations the transfer matrices do commute. Since we prove the BR
formula \eq{BRF} giving us the transfer matrices in any representation
$T_{\{\l\}}$ as a product of transfer matrices in the symmetric
representations $T_s$, we automatically establish the commutativity of
$T_{\{ \l \}}$ as expressed in \eq{COMMT}.},
\begin{equation}\label{COMMT}
    \[T_{\{\l\}}(u),T_{\{\l'\}}(u')\]=0\,.
\end{equation}
It shows that the transfer matrix, being expanded in $u$ around some point,
defines as many conserved charges in involution, as the number of degrees of
freedom of the spin chain. The Hamiltonian \eq{HEIS} is one example of a
{\it local} conserved charge since
\begin{equation*}
    \hat H_{xxx}=\left. 2\frac{d}{du }\log T(u)\right|_{u=0} \,.
\end{equation*}
with $T(u)$ corresponding to $\lambda$ being the single box fundamental representation.

A functional relation on the transfer matrices which is of special interest
for us in this paper is the determinant representation of transfer matrix in
arbitrary auxiliary $gl(K)$ irrep $\l=(\l_1,\dots,\l_a)$, $1\le a\le K-1$
\begin{equation}\label{BRF}
    T_{\{\lambda\}}(u)=\frac{1}{S_N(u)}\,\det_{1 \le i,j \le a } T_{\l_i+i-j}(u+2-2i)
\end{equation}
where we denoted by $T_s(u)$ the transfer matrices for the symmetric
representation $\lambda=1^s$ with the Young tableau given by a single row
with $s$ boxes. Notice that due to \eq{COMMT} this determinant is well
defined and there is no ambiguity concerning the order by which the
symmetric transfer matrices are multiplied. The polynomial $S(u)$ takes a
particularly simple form,
\begin{equation}\label{SPOL}
S_N(u)=\prod_{n=1}^N\prod_{k=1}^{a-1} (u-\theta_a-2k).
\end{equation}
This formula was conjectured by Bazhanov-Reshitikhin \cite{BR} (in
the absence of the twist $g$). A similar formula, with a sketch of
the proof, first appeared in the mathematical literature
\cite{Cherednik2,Cherednik3,Cherednik4}, but it is not easy to
recognize it for the physicist. In \cite{Bazhanov:2001xm} this
formula was derived for the $SL(3)$ case in the context of studies
of Conformal Field Theories with extended conformal symmetry
generated by the $W_3$ algebra\footnote{The $T$ and $Q$ operators
considered in this work would correspond from the algebraic point of
view to Baxter operators based on the quantum algebra
$U_q(\hat{sl}(3))$.}.

To shortcut these, highly abstract, mathematical constructions we propose a
more "physical", and a very direct,   proof of Bazhanov-Reshetikhin (BR)
formula,   generalizing it to an arbitrary twist $g$, as in \eq{TMATR}.
Actually, this twist appeares to be a very useful tool for the complete and
elementary proof of BR formula which we present in this paper.

More than that, our new result is the  proof of Bazhanov-Reshetikhin formula
\eq{BRF} in case of the  twisted transfer-matrices for the super-spins with
$gl(K|M)$ symmetry. To our knowledge, the super-BR formula first appeared in
\cite{Tsuboi:1997iq} and was only a
conjecture by now. In this case, the $R$-matrices $R^{\l}(u)$, where
$\l$ is a general irrep defined by a Young supertableaux
$\lambda=(\l_1,\dots,\l_a)$ (see
\cite{Kac,BahaBalantekin:1980qy,Bars1,Bars2,Bars3} for the description of
 super-irreps and
super-Young tableaux), are also known. The transfer matrix in the
supersymmetric case is defined as in \eq{TMATR}, with the super-$R$ matrices
as the entrees and with the trace replaced by a supertrace. The twist $g\to
gl(K|M)$ is a supergroup element. We will prove that the same BR formula
\eq{BRF} holds also for the twisted transfer matrices in the supersymmetric
case.

Our proof is mostly based on the properties of (super)characters. It is  a
reasonable approach since the BR formula is a natural generalization of the
Jacobi-Trudi formula for (super)characters. In this respect, the
supersymmetric case does not create much more difficulties for us than the
case of usual $gl(K)$ groups.

The BR formula  allows  to take an interesting  venue for exploiting
the integrability of quantum spin systems, rather different from the
standard coordinate or algebraic Bethe ans\"atze, to reach the
system of nested Bethe ansatz equations (BAE). Namely, the problem
of diagonalization of the transfer-matrix can be reformulated as a
problem of solving the Hirota equation describing the discrete
classical dynamics in the fusion space: the space of representations
with rectangular Young tableaux $\l=a^s$ and the spectral parameter
$u$ \cite{Klumper,Kuniba,KLWZ,Zabrodin:1996vm}. Even in the
supersymmetric case, using the "fat hook" boundary conditions in the
representation space worked out in
\cite{FD5,Tsuboi:1997iq,Tsuboi:1998sc}, it allows to obtain the
system of BAE's operating with the purely classical instruments of
integrability: B\"acklund transformations and zero curvature
representation, accompanied by the analyticity arguments \cite{KSZ}.
The supersymmetric case, being more general, allows to formulate a
new type of relations, QQ-relations \cite{KSZ}, related to the so
called fermionic duality transformations
\cite{FD1,FD2,FD3,FD4,FD5,BKSZII,KSZ} completing the TQ-relations of
Baxter, and arrive at the nested BAE's in the shortest way. A
different type of $QQ$ relations exist for nested Bethe ansatze
based on both bosonic and super algebras and are called bosonic
dualities \cite{GV,BazhanovTsuboi}.

The twisting of the super-spin chain by a group element $g\in gl(K|M)$
described above, can be naturally incorporated into this method
\cite{Zabrodin}. It is a useful tool for our derivation of BR-formula
allowing to use the nice properties of usual $gl(K|M)$-(super)characters.

\section{Transfer-matrix and BR formula in
terms of group derivatives }

We will present in this section the representation of the twisted
supersymmetric monodromy matrix and the T-matrix in terms of certain
differential operators acting  on the group. One of the advantages of this
representation will be the extensive use of characters which will appear
later to be very useful for proving various functional relations leading to
the BR formula.

The monodromy matrix $L_{\lambda}(u)$ of $N$ spins is the quantity inside
the trace in the transfer matrix \eq{TMATR},
\begin{equation}
\nn
  \hat L_{\{\l\}}\(u\) =  \[R^\l(u-\th_N) \otimes \dots
  \otimes R^\l(u-\th_1)\]  \pi_\lambda(g)\,,
\end{equation}
where the product of $R$-matrices goes along the auxiliary space with irrep
$\l$. The key idea is to rewrite this expression using a differential
operator which we call the left co-derivative,
\begin{equation}
 \left.   \hD f(g)= \frac{\d}{\d\phi}\otimes
 f(e^{\phi\cdot e}g)\right|_{\phi=0} \la{Ddef}
\end{equation}
where $\phi$ is a matrix in the fundamental representation and $\phi\cdot
e\equiv \sum_{\alpha\beta} e_{\alpha\beta} \phi^{\alpha}_{\,\beta}$.

This differential operator acts on the group element $\pi_\lambda(g)$
multiplying it by the generator $\pi_\lambda (e_{\a\b})$, present in the
$R$-matrices \eq{RMATR}, as desired. Then we can write the monodromy matrix
as
\begin{equation}\label{MONOD}
  \hat L_{\{\l\}}\(u\) =    (u_1+2 \hat D)\otimes(u_2+2 \hat D)
  \otimes(u_N+2 \hat
   D)\pi_\l(g) \,\,\, , \,\,\, u_n\equiv u-\theta_n
\end{equation}
where the matrix product of each of the $N$ factors goes along the auxiliary
space with irrep $\l$. From here we obtain the transfer matrix
$T_{\{\l\}}(u)={\rm tr}_\l L^\l\(u\)$ for which we have the following
representation in terms of left co-derivatives:
\begin{equation}\label{DIFFT}
T_{\{\l\}}(u) = (u_1+2 \hat D)\otimes(u_2+2 \hat D)
  \otimes(u_N+2 \hat
   D)\,\,\,\chi_{\{\l\}}(g)
\end{equation}
where $\chi_{\{\l\}}(g)=\tr\pi_\l(g)$ is the character of the group element
$g$ in the irrep $\l$.

The BR formula  \eq{BRF}  in our notations claims that
\begin{equation}\label{BRFTL}
T_{BR}^{\{\lambda\}}(u)= S(u)\,T_{\{\lambda\}}(u)
\end{equation}
where
\begin{equation}\label{BRFT}
T_{BR}^{\{\lambda\}}(u)\equiv \det_{1 \le i,j \le a }\[ (u_1+2-2i+2
\hD)\otimes\dots\otimes(u_N+2-2i+2 \hD)\chi_{\l_i+i-j}(g)\],
\end{equation}
$\chi_{s}(g)$ is the character of symmetric irrep $1^s$ (Schur polynomial)
and $T_{\{\lambda\}}(u)$ is the transfer matrix. The polynomial $S(u)$ is
given by \eq{SPOL}. In the next two sections we prove this main statement.

In the rest of this section we will precise the meaning of  the left
co-derivative defined through \eq{Ddef} and give some useful formulas for
it. Let us recall that $\hD$ carries only the indices of the quantum spin
and that $\frac{\d}{\d\phi}$ is a matrix derivative. Explicitly we have
\begin{equation}
\frac{\partial}{\partial \phi_{i_1}^{\,j_1}}\phi_{\,j_2}^{i_2} =
\delta_{j_1}^{i_2} \delta_{j_2}^{i_1} \la{dd} \,.
\end{equation}
and thus
\begin{equation*}
 \hat D_{j_1}^{i_1}
g_{j_2}^{i_2}=\left.\frac{\partial}{\partial \phi^{\,j_1}_{i_1} } \(
e^{\phi\cdot e }g \)_{j_2}^{i_2} \right|_{\phi=0}= \delta_{j_1}^{i_2}
g_{j_2}^{i_1} \,
\end{equation*}
so that we see that the "out-going" indices $j_1,j_2$ are untouched whereas
the "in-coming" indices $i_1,i_2$ are swapped. We can thus write the
previous relation in a more abstract and elegant way as
\begin{equation}
\frac{\d}{\d\phi} \otimes \phi = \mathcal{P} \,\,\, , \,\,\, \hat D \otimes
g= \mathcal{P} \,(1\otimes g) \,,\nn
\end{equation}
  where $\mathcal{P}$ is the permutatation operator defined through
$\mathcal{P} \,(A \otimes B)= (B \otimes A) \, \mathcal{P}$. This formula
easily generalizes to
\begin{equation}
 \hat D \otimes g^n=\sum_{k=0}^{n-1} \mathcal{P}\,(g^k\otimes
 g^{n-k})\,. \la{gn}
\end{equation}
Let us also write down the following formulas useful for the future
\begin{eqnarray}\label{USEFUL}
\hD \,{\rm tr}\,\log(1-g z)&=& \frac{gz}{1-g z}\,,\nn\\
 \hD \otimes \frac{gz}{1-g z}&=& \mathcal{P}\,\(\frac{1}{1-g z}\otimes
  \frac{g z}{1-g z}\)\,.
\end{eqnarray}
Obviously, they follow from \eq{gn}.

\section{The proof of the one spin BR formula}\la{sec1spin}

In this section, to demonstrate the idea of the proof,  we will consider a
simpler case of the single site spin chain. Many features of the full proof
of the BR formula are contained already in this example. The BR formula
\eq{BRF} in this one spin case claims that
\begin{equation}\nn
T_{BR}^{\{\lambda\}}(u)\equiv \det_{1 \le i,j \le a }\[ (u+2-2i+2
\hD)\chi_{\l_i+i-j}(g)\]
\end{equation}
equals
\begin{equation}\nn
S_1(u)\,T_{\{\lambda\}}(u)=\prod_{k=1}^{a-1}(u-2k) \,(u+2\hat
D)\chi_{\l}(g)\,.
\end{equation}
Our strategy of the proof  will be as follows: we start by proving that the
$a^{th}$ order polynomial $T_{BR}^{\{\lambda\}}(u)$ has indeed zeroes
precisely at $u=2,4,\dots,2a-2$. Having done this we can read off the
remaining (linear) factor from the large $u$ asymptotics. We will see that
it matches precisely $T_{\{\lambda\}}(u)$.

Indeed, let us put  in \eq{BRF} successively $u=2k,\,\,\, k=1,2,\dots,
2(a-1)$ and look at two neighboring $k$-th and $(k+1)$-th columns of the
matrix under the determinant,
\begin{equation}
T_{BR}^{\{\l\}}(2k)=\(\begin{array}{cllc}
\dots & T_{\l_k+k-1}(2) & T_{\l_{k+1}+k}(0) & \dots \\
\dots & T_{\l_k+k-2}(2) & T_{\l_{k+1}+k-1}(0) & \dots \\
\dots & \dots & \dots & \dots  \\
\dots &T_{\l_k+k-a+1}(2) & T_{\l_{k+1}+k-a+2}(0) & \dots \\
\dots &T_{\l_k+k-a}(2) & T_{\l_{k+1}+k-a+1}(0) & \dots
  \end{array}\) \,.
\end{equation}
Any $2\times 2$ minor of the sub-matrix formed from these two columns is of
the form
\begin{equation}\label{MINOR}
T_{s_1}\(2\)  T_{s_2}\(0\)- T_{s_1+1}\(2\) T_{s_2-1}\(0\)=
4\[(1+\hD)\chi_{s_1}\] \cdot\[ \hD\chi_{s_2}\]-4\[(1+\hD)\chi_{s_1+1}\]
\cdot\[  \hD\chi_{s_2-1}\]
\end{equation}
We will show that any such minor, and therefore the whole determinant $
T_{BR}^{\{\lambda\}}(2k)$ is zero, which proves the statement about the
positions of zeroes. Let us remind that the left co-derivatives $\hat D$ act
here only on the next following character, whereas the terms in square
brackets are multiplied as matrices  in quantum space of the single spin.

To prove this identity we use the  generating function of the characters
$\chi_s$ in symmetric  irreps (Schur polynomials),
\begin{equation}\label{DEFW}
    w(z)\equiv {\rm det}\, \(1-z g\)^{-1}
    =\sum_{s=1}^\infty \chi_s \,z^s \,.
\end{equation}
The identity (\ref{MINOR}) is a trivial consequence of
\begin{equation}\label{ONESPINID}
   \(1+\hD\) w(z_1) \cdot \hD\,w(z_2)=
 \hD \frac{w(z_1)}{z_1}\cdot \(1+\hD\) z_2\,w(z_2)
\end{equation}
which follows immediately from the first of  \eq{USEFUL}. Thus we proved
that the BR transfer matrix \eq{BRF} is indeed given by a trivial factor
$S(u)$ times some operator linear in $u$. To read off this operator we
expand
\begin{equation}
\frac{1}{S_1(u)}\,T_{BR}^{\{\lambda\}}(u)=u \det_{1 \le i,j \le a }\[
\(1+\frac{2}{u+2-2i} \hD\)\chi_{\l_i+i-j}(g)\] \nn
\end{equation}
 at large $u$ to find
\begin{eqnarray}\label{BRFPR}
    \frac{1}{S_1(u)}\hat T_{BR}^{\{\lambda\}}(u)
    &\to&u \det_{1\le i,j\le a}\chi_{\l_j+i-j} +
    2\sum_{k=1}^a \det_{1 \le i,j \le a }\[ \((1-\delta_{j,k})+ \delta_{j,k}\hD\)
    \chi_{\l_i+i-j}\] \,\, , \,\, u \to \infty\nn\\
    &=&(u+2 \hat D) \chi_\l\nn
\end{eqnarray}
where we have used the Jacobi-Trudi formula\footnote{In section
\ref{supersec} we shall explain how to generalize all derivations for the
superalgebras $gl(K|M)$. In this case the Jacobi-Trudi formula still holds
and moreover $a$ can take any positive integer value, provided that the "fat
hook" condition $\l_{K+1}\le M$ is satisfied.} for the $gl(K)$-character in
the irrep $\lambda$ (see the Appendix A for its demonstration)
\begin{eqnarray}\label{CHADET}
 \chi_{\{\l\}}(g) = \det_{1\le i,j\le a}\chi_{\l_j+i-j} (g)\,.
\end{eqnarray}
Hence we proved the BR formula for one spin.

\section{The proof of the full multi-spin BR formula} \la{secgen}

Here we will generalize our proof to the general $N$-spin BR formula.
Namely, we will show that the BR determinant representation of  $T$-matrix
\eq{BRF} is equivalent to  the original definition of the transfer-matrix
\eq{DIFFT}.

First of all let us reduce the proof of the BR formula to the proof of the
identity
\begin{equation}\label{SPINIDgen}
 \[  (1+\hD)^{ \otimes N}  \, w(z_1) \] \cdot \[ \hD^{\otimes N} \,w(z_2)\]=
 \[\hD^{\otimes N} \, \frac{w(z_1)}{z_1}\] \cdot \[ (1+\hD)^{\otimes N}
 \,z_2\,w(z_2) \]     \,,
\end{equation}
generalizing \eq{ONESPINID}. This identity will be proved in the next
section.

The logic goes as for the single spin case. We start by showing that the
operator
\begin{equation}\label{BRFgen}
T_{BR}^{\{\l\} }= \det_{1 \le i,j \le a }   T_{s_i+i-j}(u+2-2i) \,,
\end{equation}
contains the trivial factor
\begin{equation}
S_N(u)=\prod_{n=1}^{N} \prod_{k=1}^{a-1} (u-\theta_n-2k) \,. \nn
\end{equation}
As before -- see \eq{MINOR} -- this follows from
\begin{equation}
T_{s_1}\(\theta_n+2\) T_{s_2}\(\theta_n\)-
T_{s_1+1}\(\theta_n+2\)T_{s_2-1}\(\theta_n\)=0 \la{TTgen}
\end{equation}
which turns out to be equivalent to \eq{SPINIDgen} as we shall now explain.
Indeed suppose \eq{TTgen} is true for a spin chain of length $N$ and suppose
we want to check it for $N+1$ spins at $u=2+\theta_n$. We write it as
\begin{eqnarray*}
\label{LU}
  0 &\stackrel{?}{=}&\(\theta_n-\theta_{N+1}+2+2\hD \) \otimes  L_{s_1,N}(2+\theta_n) \cdot
  \(\theta_n-\theta_{N+1}+2\hD \) \otimes  L_{s_2,N}(2+\theta_n) \\
   && -   \(\theta_n-\theta_{N+1}+2+2\hD \) \otimes  L_{s_1+1,N}(2+\theta_n) \cdot \(\theta_n-\theta_{N+1}+2\hD \)
   \otimes  L_{s_2-1,N}(2+\theta_n)
 \end{eqnarray*}
and we see that the $\theta_n-\theta_{N+1}$ dependent terms are proportional
either to the identity with $N$ spins or to the derivative of this identity!
Thus, to check this relation we can set $\theta_{N+1}=\theta_n$. Repeating
this procedure for every $n$, and knowing that the identity is true for
$N=1$, we conclude, by induction, that to check the identity \eq{TTgen} it
suffices indeed to prove \eq{SPINIDgen}. This main identity will be proven
in the next section. In the remaining of this section let us take it as
granted and finish the proof of the BR formula.

Having identified the trivial factor $S(u)$ inside $T_{BR}^{\{\l\}}(u)$ we
write it as
\begin{equation}\nn
\frac{1}{S(u)} T_{BR}^{\{\l\} }=u_1u_2\dots u_N \det_{1 \le i,j \le a }
\,\bigotimes_{n=1}^{N} \(1 +\frac{2}{u_n+2-2i} \hD \) \chi_{s_i+i-j} \,,
\end{equation}
where $u_n=u-\theta_n$. We know that the r.h.s. must be a linear polynomial
in each of the variables $u_n$. We can then read this polynomial from the
large $u_n$ asymptotics. For example, for large $u_1$ we find
\begin{equation}\label{BRFgen1spin}
\frac{1}{S(u)} T_{BR}^{\{\l\} }\to u_2\dots u_{N} \(u_1+2 \hD \) \det_{1 \le
i,j \le a }   \,\bigotimes_{n=2}^{N} \(1 +\frac{2}{u_n+2-2i} \hD \)
\chi_{s_i+i-j} \,.
\end{equation}
Expanding in this way for each of the remaining $u_n$'s we clearly recover
\begin{equation}
T_{\{\lambda\}}(u)=\bigotimes_{n=1}^{N} (u_n+2 \hD) \,\chi_{\lambda}
\end{equation}
and thus prove the BR conjecture. In the next section we will fill the gap
in this derivation by proving the general identity \eq{SPINIDgen}.
\subsection{The general identity}
\begin{figure}[h]    \centering
        \resizebox{151mm}{!}{\includegraphics{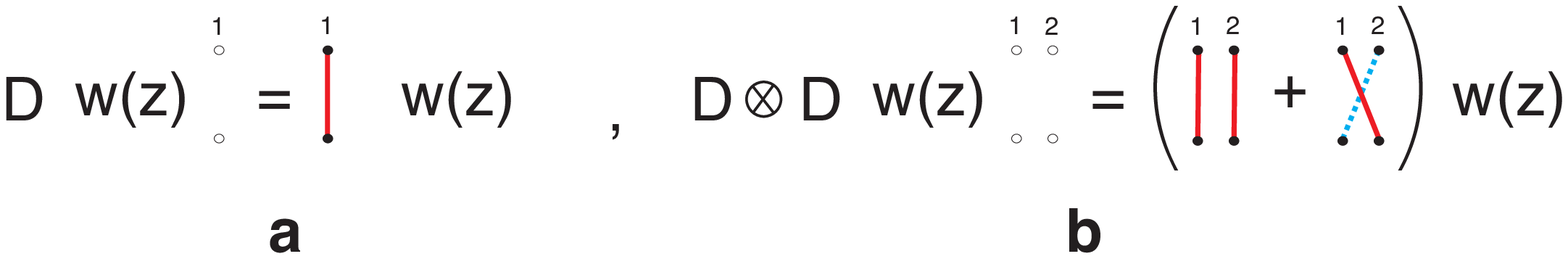}}
   \caption{ \footnotesize  A bold solid line from upper node $n$
   to lower node $m$ represents a $\(\frac{gz}{1-gz}\)^{i_n}_{j_m}$
   factor whereas a dashed line corresponds to $\(\frac{1}{1-gz}\)^{i_n}_{j_m}$.
    In figure \ref{fig:BR1}a we represent the action of the left co-derivative
    on the symmetric generating function. In figure \ref{fig:BR1}b we add an extra
    derivative (living on a new quantum space represented by the empty
    balls at position $2$) which will yield two new terms corresponding
     to the action on the generating function and on the previously created line.
   }\label{fig:BR1}
\end{figure}

\la{diagramatics}
In this section we shall prove the identity \eq{SPINIDgen} which was the key
ingredient in the proof of the BR formula. To do so we need to understand in
great detail the objects involved in this identity, namely
\begin{equation}
\hD^{\otimes N} w(z) \,\,\, \text{and} \,\,\, (1+\hD)^{\otimes N} w(z) \,,
\end{equation}
where $w(z)$ is the generating function \eq{DEFW} introduced above. From
\eq{USEFUL} we have
\begin{equation}
\hD  \,w(z) =\frac{gz}{1-gz}w(z)\,\,\, \Leftrightarrow \,\,\, \[\hD
w(z)\]_{j_1}^{i_1} =\[\frac{gz}{1-gz} \]_{j_1}^{i_1} w(z) \,. \la{1spins}
\end{equation}
This relation can be represented graphically as in figure \ref{fig:BR1}a
using a solid line from an upper to a lower point to indicate the term in
brackets in this equation.

If we act on this expression with a second left co-derivative (in a new
quantum space) we get, using \eq{USEFUL} again,
\begin{equation}
\hD\otimes \hD  \,w(z) =\[\frac{gz}{1-gz}\otimes
\frac{gz}{1-gz}+\mathcal{P}\, \frac{1}{1-gz}\otimes \frac{gz}{1-gz} \]w(z)
\end{equation}
where the first term comes from the derivative acting of the generating
function $w(z)$ while the second term comes from the action of the
derivative on the factor $gz/(1-gz)$. If we want to make the indices
manifest this is the same as
\begin{equation}
\[\hD\otimes \hD  \,w(z) \]^{i_1i_2}_{j_1j_2}=\[\(\frac{gz}{1-gz}\)^{i_1}_{j_1}
 \(\frac{gz}{1-gz}\)^{i_2}_{j_2}+\( \frac{1}{1-gz} \)^{i_2}_{j_1}
  \( \frac{gz}{1-gz} \)^{i_1}_{j_2} \]w(z) \,. \la{2spins}
\end{equation}
In the second term the permutation operator swaps the "in-coming" indices
$i_1,i_2$. We will often denote the upper indices $i_a$ by "in-coming" and
the lower indices $j_a$ by "out-going". This relation is graphically
presented in figure \ref{fig:BR1}b.

\begin{figure}[h]    \centering
        \resizebox{151mm}{!}{\includegraphics{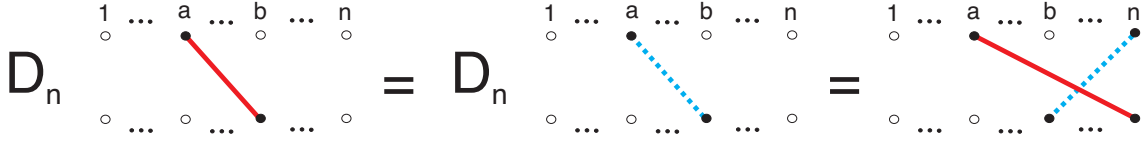}}
   \caption{ \footnotesize The action of the left co-derivative, living
   on a new quantum space $n$,  on a line going from upper $a$ to lower
    $b$ positions generates a \textit{dashed} line going to left from upper
     position $n$ to lower position $b$ and a \textit{solid} line going to
     the right from the upper position $a$ to the lower position $n$. The final result is independent on whether the original line is dashed or solid.
   }\label{fig:BR2}
\end{figure}

Suppose we now act by a third derivative $\hD$ with new indices in a third
quantum space. It can either hit the generating function $w(z)$, yielding a
factor $ \(\frac{gz}{1-gz}\)^{i_3}_{j_3}$, drawn as a vertical solid line,
or it can act on one of the factors $\(\frac{gz}{1-gz}\)^{i_a}_{j_b}$ or
$\(\frac{1}{1-gz}\)^{i_a}_{j_b}$ in \eq{2spins}. The action of the
derivative on these factors, depicted as a  solid or  dashed line
respectively, is the same regardless of the presence or absence of the
factor $gz$ in the numerator because these factors differ by $1$,
\begin{equation}
\hD^{i_n}_{j_n}  \(\frac{gz}{1-gz}\)^{i_a}_{j_b}=\hD^{i_n}_{j_n}
\(\frac{1}{1-gz}\)^{i_a}_{j_b}=\(\frac{1}{1-gz}\)^{i_n}_{j_b}
\(\frac{gz}{1-gz}\)^{i_a}_{j_n}
\end{equation}
which is represented graphically as in figure \ref{fig:BR2}.

We see that when we add an extra derivative on a new quantum space $n$ this
derivative acts on any line going from upper position $a$ to lower position
$b$ creating two new lines: One going to the \textit{left} from upper
position $n$ to lower position $b$ and another one, going to the
\textit{right} from the upper position $a$ to the lower position $n$. This
does \textit{not} depend on the nature of the original line going from upper
$a$ to lower $b$, that is whether it is a dashed or a solid line. Notice
furthermore that, of the two generated lines, the one going to the right is
always solid whereas the one going to the left is always dashed.

\begin{figure} [h]   \centering
        \resizebox{151mm}{!}{\includegraphics{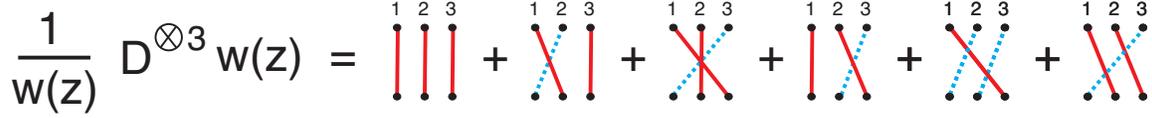}}
   \caption{ \footnotesize To compute
   $\[\hD^{\otimes N} w(z)\]^{i_1\dots i_N}_{j_1\dots j_N}$
    we draw all $N!$ permutation diagrams, dash the lines going
    to the left and read the contribution of each term from the
     rule that a dashed (solid) line going from upper position $a$
      to lower position $b$ represents a factor
      $\(\frac{1}{1-gz}\)^{i_a}_{j_b}$ $\(\(\frac{gz}{1-gz}\)^{i_a}_{j_b}\)$
      respectively.   }\label{fig:BR3}
\end{figure}

It is clear how the action of $N$ derivatives on $w(z)$ will look like -- we
will get the $N!$ possible permutation diagrams with dashed or solid lines
connecting the "in-coming" and "out-going" indices. Vertical lines are
generated only when the left co-derivative acts on $w(z)$ and should thus
always be solid. The lines going to the left and right are created when the
differential operator acts on some already created line as described above.
Thus lines going to the right (left) are always solid (dashed).  In figure
\ref{fig:BR3} we represent the action of $\hD\otimes \hD\otimes \hD$ on the
generating function $w(z)$.

Algebraically this can be summarized as
\begin{equation}
\[\hD^{\otimes N} w(z)\]^{i_1\dots i_N}_{j_1\dots j_N}=
\sum_{\text{Permutations} \,P}^{N!} \prod_{k=1}^N \(
\frac{(gz)^{\theta(P_k-k)}}{1-gz} \)^{i_k}_{j_{P_k}} \la{dw}
\end{equation}
where the Heaviside theta function vanishes for lines going to the left
($P_k<k$) and equals one for vertical and right-going lines ($P_k\ge k$) in
accordance with the  rules described above.

\begin{figure} [h]   \centering
        \resizebox{151mm}{!}{\includegraphics{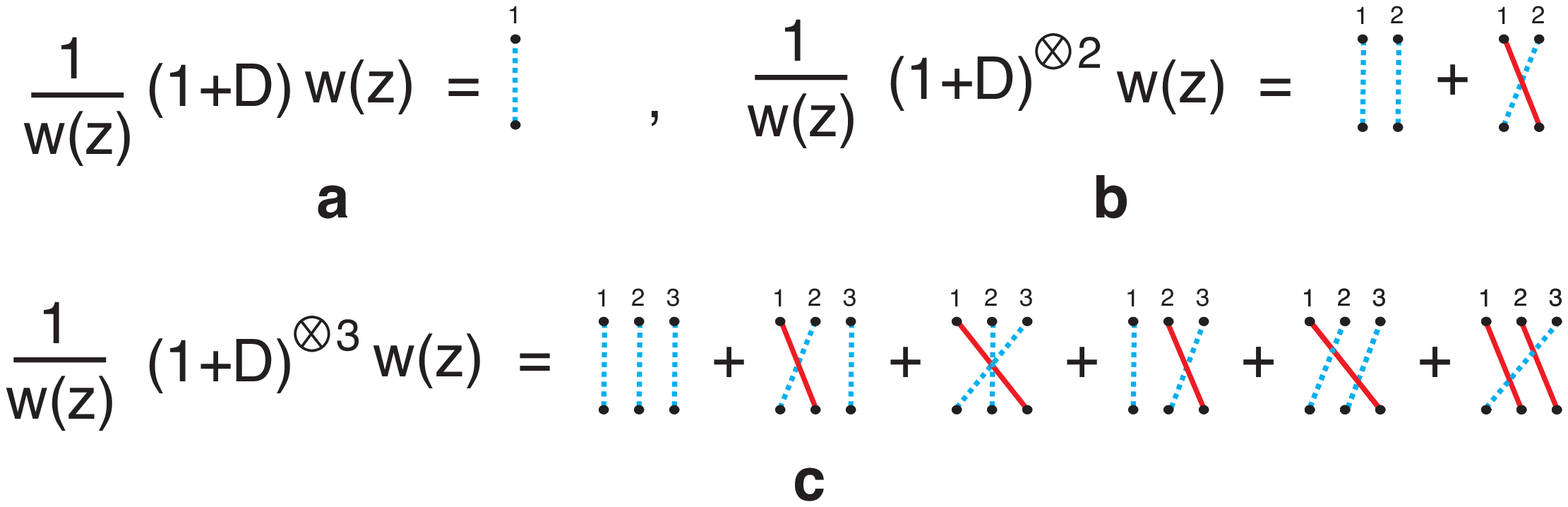}}
   \caption{ \footnotesize To compute
   $\[(1+\hD)^{\otimes N} w(z)\]^{i_1\dots i_N}_{j_1\dots j_N}$
   we draw all $L!$ permutation diagrams, dash the vertical and
   left-going lines and read the contribution of each term from
   the rule that a dashed (solid) line going from upper position
   $a$ to lower position $b$ represents a factor
   $\(\frac{1}{1-gz}\)^{i_a}_{j_b}$ $\(\(\frac{gz}{1-gz}\)^{i_a}_{j_b}\)$
   respectively. In figures \ref{fig:BR4}a,b,c we represent the outcome for $N=1,2,3$.  }
   \label{fig:BR4}
\end{figure}

Having understood what $\hD^{\otimes N} w(z)$ is, let us consider the other
object appearing in \eq{SPINIDgen}, namely
\begin{equation}
(1+\hD)^{\otimes N} w(z) \,.
\end{equation}
In fact this object is also given by an equally simple set of graphical
rules. For $N=1$
\begin{equation}
(1+\hD)  w(z) = \frac{1}{1-gz} w(z) \,\,\, \Leftrightarrow \,\,\,
\[(1+\hD)  w(z) \]^{i_1}_{j_1}= \(\frac{1}{1-gz}\)^{i_1}_{j_1}
w(z)\,, \la{previous}
\end{equation}
which in our graphical rules corresponds to a dashed vertical line as in
figure \ref{fig:BR3}a. Next we take $N=2$. That is, we apply the operator
$(1+\hD)$ (with open indices living in a new quantum space) to the previous
expression. The trivial $1$ in this operator just translates into a
Kronecker delta function $\delta^{i_2}_{j_2}$ multiplied by the previous
expression \eq{previous}. The derivative then yields two type of terms: If
it hits the generating function it simply produces a factor of
$\(\frac{gz}{1-gz}\)^{i_2}_{j_2}$ which again multiplies the previous
expression \eq{previous}; If it acts on the $\(\frac{1}{1-gz}\)^{i_1}_{j_1}$
factor it will create two lines as depicted in figure \ref{fig:BR2} thus
giving rise to a different permutation diagram. Thus the contribution from
the $1$ can be combined with the contribution coming from the action of the
left co-derivative $\hD$ on the generation function $w(z)$ to transform the
factor of $\(\frac{gz}{1-gz}\)^{i_2}_{j_2}$ into
$\(\frac{1}{1-gz}\)^{i_2}_{j_2}$ as
\begin{equation}
1+\frac{gz}{1-gz}=\frac{1}{1-gz}\,.
\end{equation}
In total, for $N=2$ we get
\begin{equation}
(1+\hD)^{\otimes 2}  w(z) = \[\frac{1}{1-gz}  \otimes \frac{1}{1-gz}
+\mathcal{P} \,\frac{1}{1-gz}  \otimes \frac{g}{1-gz} \] w(z)  \,,
\end{equation}
which we represent in figure \ref{fig:BR4}b. The three spin case is depicted
in figure \ref{fig:BR4}c.

For a generic number of spins $N$ the pattern should now be obvious. The
presence of the one in the operator $(1+\hD)$ simply makes the vertical
lines -- comming from the action of the derivative $\hD$ on the generating
function -- dashed instead of solid as before. That is to compute
$(1+\hD)^{\otimes N}w(z)$ we simply sum all the $N!$ permutation diagrams
where lines going to the right from upper "in-coming" indices to lower
"outgoing" indices are solid whereas vertical and left-going lines are
dashed. Algebraically,
\begin{equation}
\[(1+\hD)^{\otimes N} w(z)\]^{i_1\dots i_N}_{j_1\dots j_N}=
\sum_{\text{Permutations} \,P}^{N!} \prod_{k=1}^N \(
\frac{(gz)^{\theta(P_k-k-1)}}{1-gz} \)^{i_k}_{j_{P_k}} \la{1dw}
\end{equation}

Given the strikingly similarity between these  two objects,
$(1+\hD)^{\otimes N} w(z)$ and $\hD^{\otimes N} w(z)$, it is natural to
expect  some simple relation between them which we will establish now.

\begin{figure} [h]   \centering
        \resizebox{151mm}{!}{\includegraphics{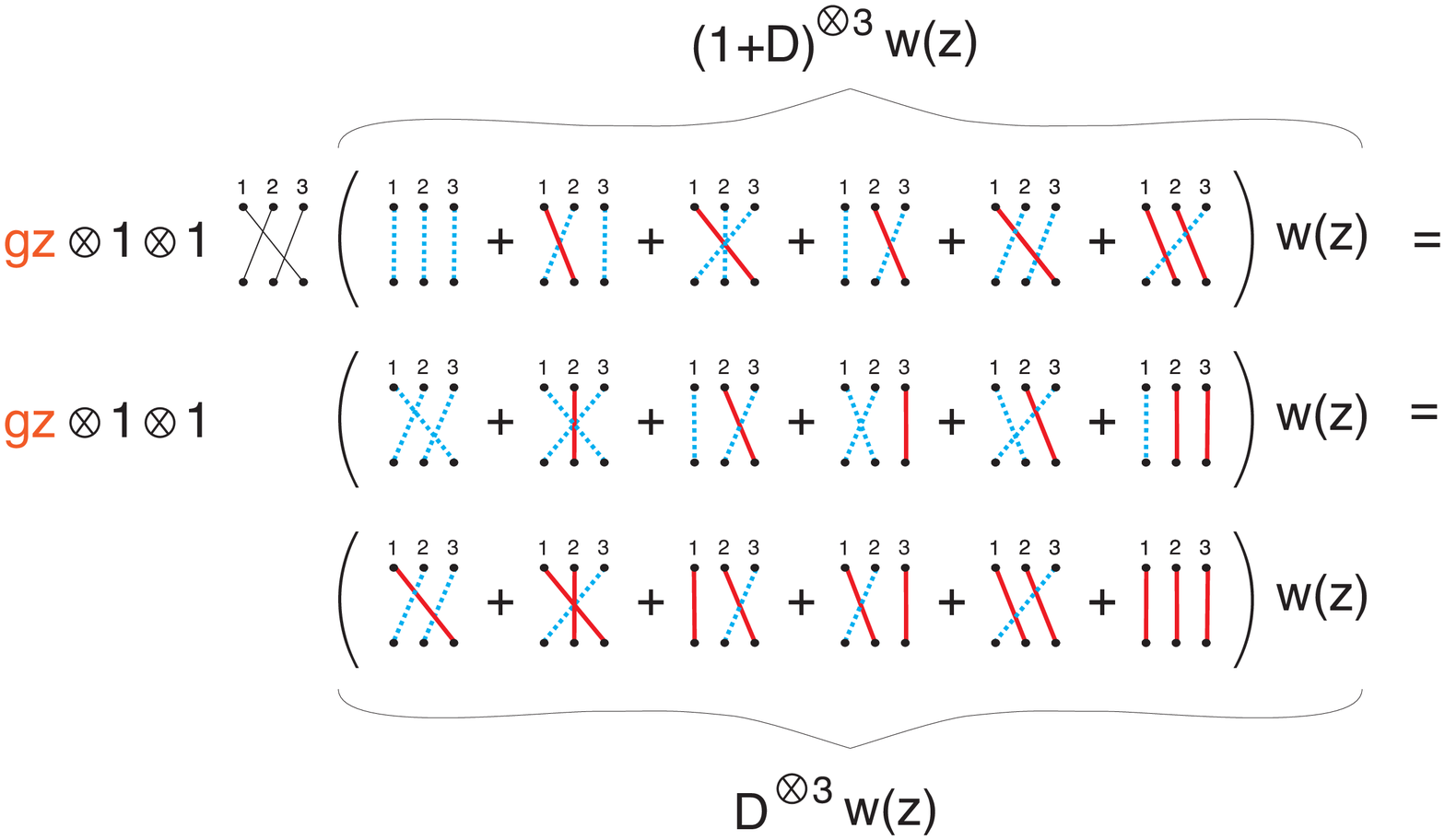}}
   \caption{ \footnotesize Illustration of the identity \eq{id1} for $N=3$.
    In the first and last line we can see the matrices
    $(1+\hD)^{\otimes 3}w(z)$ and $\hD^{\otimes 3}w(z)$ respectively.
    The graphical rules for these objects differ only for the vertical
     lines which should be dashed for $(1+\hD)^{\otimes 3}w(z)$ and solid
      for  $\hD^{\otimes 3}w(z)$. When we apply the shift operator,
      represented  in the first line by a set of thin (black) lines ,
      to $(1+\hD)^{\otimes 3}w(z)$ the new vertical lines come from
      previously right-tilted lines and will thus be solid lines as they
      should for the $\hD^{\otimes N} w(z)$ operator. The only lines which
       are  wrong in the set of graphs after the first equal sign are
        those starting from the first "in-coming" index because these
        lines are the image of the lines which started on the last "in-coming"
         index in  $(1+\hD)^{\otimes 3}w(z)$ and which were, therefore, dashed.
          The factor $g z\otimes \one \otimes \one$ should be there to correct
          this first line making it always solid as it should be for the object
          $\hD^{\otimes 3}w(z)$.}\label{fig:BR5}
\end{figure}

Indeed suppose we shift all "in-coming" indices $i_a$ in $(1+\hD)^{\otimes
N} w(z)$ to the right,
\begin{eqnarray*}
i_1 &\to& i_2 \\
i_2&\to & i_3 \\
&\dots& \\
i_N&\to& i_1\,.
\end{eqnarray*}
In other words we multiply $(1+ \hD)^N w(z)$  by the cyclic shift operator
$U_L\equiv \mathcal{P}_{12}\mathcal{P}_{23}\dots \mathcal{P}_{L-1,L}$.
 For now let us
ignore the lines originally starting at the last incoming index $i_N$. After
the application of the shift operator, lines which were going to the left
from the upper to the lower indices are now even more tilted and of course
still go to the left. Lines which go to the right with a large tilt will
still go to the right but with a smaller inclination. An interesting
phenomenon happens then for vertical and for minimally tilted ($i_a$ united
with $j_{a+1}$) lines going to the right. Vertical lines -- which were
dashed lines -- will become left-going dashed lines whereas the minimally
tilted right-going lines -- which were solid lines -- will become vertical
solid lines. Thus after application of the twist operator the vertical and
right-going lines are solid and the left-going lines are dashed. But these
are precisely the graphical rules for $\hD^{\otimes N} w(z)$! Finally let us
consider the lines starting at the last incoming index $i_N$ which we
ignored so far. In $(1+\hD)^N w(z)$ the lines starting from this point must
always be dashed because they can only go to the left or be vertical. Under
the application of the twist operator this index becomes the first
"in-coming" index $i_1$. In $\hD^N w(z)$ lines leaving this first "in-coming
point" should always be solid because they are either vertical or go to the
right. Thus, if we want to relate the shifted $(1+\hD)^{\otimes N}$ with
$\hD^N$ the only correction we should make is to transform the dashed line
leaving the first "in-coming" index in the twisted $(1+\hD)^{\otimes N}$
into a solid line. This can be trivially made by multiplication of
$gz\otimes \one \otimes \dots \otimes \one$, that is
\begin{equation}
\hD^{N} w(z)=\[\(g \otimes \one \otimes \dots \otimes \one\) U_L \]
\,(1+\hD)^{\otimes N} z w(z) \,. \la{id1}
\end{equation}
In figure \ref{fig:BR5} this identity is exemplified on the three spin case.
Notice also that using the explicit expressions \eq{dw} and \eq{1dw} this
identity can be checked through a straightforward algebraic computation.

Following exactly the same kind of reasonings we can also prove
\begin{equation}
(1+\hD)^{N} w(z')=\hD^{\otimes N} \,\frac{w(z')}{z'} \[\(g \otimes \one
\otimes \dots \otimes \one\) U_L \]^{-1}\,. \la{id2}
\end{equation}
Then multiplying (\ref{id2}) and (\ref{id1}) we obtain precisely the
identity (\ref{SPINIDgen}) which we aimed at.

\section{Generalization to supergroups} \la{supersec}

In this section we will explain how to generalize the derivations in the
previous sections when the symmetry group is given by the superalgebra
$GL(K|M)$. In this case the rational $R$-matrix is given by \cite{Kulish}
\begin{equation}
    R_{\l}(u)=u+2\sum_{\a\b}(-)^{p_\a}e_{\b\a}  \otimes  \pi_\l(e_{\a\b}) \,, \la{RMATRsup}
\end{equation}
where the index $\alpha$ is called bosonic ($p_\a=0$) for $1 \le \a \le K$
and fermionic ($p_\a=1$) for $K < \a \le K+M$. In the fundamental
representation the second term
\begin{equation}
\mathcal{P}\equiv\sum_{\a\b}(-)^{p_\a}e_{\b\a}  \otimes  e_{\a\b} \,,
\end{equation}
becomes the superpermutation operator. Indeed,  consider the
standard base $e_\a$ for the quantum states, such that $e_{\a\b}
e_{\delta}=\delta_{\b\delta} e_\a$. Then the action of the
superpermutator on $e_\gamma \otimes e_\delta$ can be computed by
simply moving  the basis vectors and the generators towards each
other, everyone to its space, adding a braiding factor $(-1)^{p_\a
p_\b}$ whenever an index $\a$ is moved past an index $\beta$. It is
then simple to check that
\begin{eqnarray*}
\mathcal{P}_{12} \, e_\gamma \otimes  e_\delta &=& (-1)^{p_\delta p_\gamma}
e_{\delta}\otimes  e_{\gamma}\,,
\end{eqnarray*}
as expected for a superpermutation operator. Notice that the action on even
states where fermionic (bosonic) basis vectors are always contracted with
fermionic (bosonic) components,
\begin{equation}
x\equiv x^\a e_\a \,\, , \,\, y\equiv y^\a e_\a \,, \la{bosonic}
\end{equation}
is given simply by an exchange of states,
\begin{eqnarray*}
\mathcal{P}_{12} \,\( x  \otimes  y \) = y \otimes  x \,  .
\end{eqnarray*}
The action of $\mathcal{P}_{13}=\sum_{\a\b }(-)^{p_\a}  e_{\b\a}  \otimes
\one\otimes   e_{\a\b} $ on a three spin state, for example, reads
\begin{eqnarray*}
\mathcal{P}_{13} \,\( x \otimes y  \otimes z\)  = z \otimes  y   \otimes   x
\,.
\end{eqnarray*}
so that the superpermution operator $\mathcal{P}_{ij}$ simply exchanges the
even states at positions $V_i$ and $V_j$. Acting on basis vector we will
obtain the obvious additional minus signs, for example,
\begin{eqnarray*}
\mathcal{P}_{13}   \,e_\gamma \otimes  e_\delta \otimes e_\rho=(-)^{
p_\gamma p_\delta+p_\gamma p_\rho+p_\rho p_\delta}\,e_{\rho} \otimes
e_\delta  \otimes   e_{\gamma}      \,.
\end{eqnarray*}
Notice that  due to the presence of the intermediate vector
$e_\delta$ the minus sign involved when $e_\rho \leftrightarrow
e_\gamma$ is \textit{not} simply $(-1)^{p_\rho p_\gamma}$. Notice
also that
\begin{eqnarray}
\mathcal{P}_{13}=
\mathcal{P}_{12}\mathcal{P}_{23}\mathcal{P}_{12}=\mathcal{P}_{23}\mathcal{P}_{12}\mathcal{P}_{23}
\,, \la{13etc}
\end{eqnarray}
without any minus signs involved, just like for the usual permutation
operator.

The second step in our construction is to re-write the super monodromy
matrix
\begin{eqnarray}
\nn
  \hat L_{\{\l\}}\(u\) &=&  \[R^\l(u-\th_N) \otimes \dots
  \otimes R^\l(u-\th_1)\]  \otimes\pi_\lambda(g)\,.
\end{eqnarray}
as
\begin{eqnarray}\label{Lsup}
  \hat L_{\{\l\}}\(u\) &=&    (u_1+2 \hat D)\otimes(u_2+2 \hat D)
  \otimes(u_N+2 \hat
   D) \otimes\pi_\l(g)
\end{eqnarray}
where for supergroups our left co-derivative acts as follows
\begin{equation}
\hat D \otimes f(g)=e_{ij}  \frac{\partial}{\partial \phi_i^{\,j}} \otimes
f\(e^{ \phi^k_{\,l} e_{kl}} g\)_{\phi=0}\,\,\, , \,\,\,
\frac{\partial}{\partial \phi_{i_1}^{\,j_1}}\phi_{\,j_2}^{i_2} \equiv
\delta_{j_1}^{i_2} \delta_{j_2}^{i_1} (-1)^{p_{j_1}} \la{dsuper} \,.
\end{equation}
It is  instructive to check the last factor in \eq{Lsup},
\begin{eqnarray*}(u\,\one\otimes \one_\l+2   \hat D \otimes \one_\l)\otimes \pi_\l(g)&=&\left.u\, \one \otimes \pi_{\lambda}(g)+ 2 \,e_{ij}  \frac{\partial}{\partial \phi_i^{\,j}} \otimes \pi_\l \(e^{e_{kl} \phi^k_{\,l}} g\)\right|_{\phi=0} \\
&=&\left.u \,\one \otimes \pi_{\lambda}(g)+ 2\, e_{ij}  \frac{\partial}{\partial \phi_i^{\,j}} \otimes \phi^k_{\,l} \pi_\l \(e_{kl}  g\)\right|_{\phi=0} \\
&=&\[u\, \one \otimes \one_\l +2 \, (-1)^{p_{j}} e_{ij}  \otimes  \pi_\l
\(e_{ji}\) \] \pi_\l\(  g\)
\end{eqnarray*}
which is indeed precisely what one needs -- see \eq{RMATRsup}. Formulae
\eq{USEFUL} are also trivially generalized to
\begin{eqnarray}\label{USEFULsup}
\hD \,{\rm str}\,\log(1-g z)&=& \frac{gz}{1-g z}\,,\nn\\
 \hD \otimes \frac{gz}{1-g z}&=& \mathcal{P}\,\(\frac{1}{1-g z}\otimes
  \frac{g z}{1-g z}\)\,.
\end{eqnarray}
where $\mathcal{P}$ is the superpermutation and $\rm str$ is the supertrace,
${\rm str} A\equiv \sum A_{ii} (-)^{p_i}$. Since for supercharacters the
generating function \eq{DEFW} becomes
\begin{equation}\label{DEFWsup}
    w(z)\equiv {\rm sdet}\, \(1-z g\)^{-1}
    =\sum_{s=1}^\infty \chi_s \,z^s \,,
\end{equation}
and $\sdet A=\exp {\rm str} \log A$, the prove of the single spin BR
formula goes exactly as for the usual algebras $GL(K)$ in section
\ref{sec1spin}. For  many spin case, exactly as for the bosonic case
in section \ref{secgen}, we only need  to prove the identity
\eq{SPINIDgen} for the generating function of the "symmetric"
super-characters (corresponding to the one-row Young tables),
\begin{equation}
\[  (1+\hD)^{ \otimes N}  \, w(z_1) \] \cdot \[ \hD^{\otimes N} \,w(z_2)\]=
 \[\hD^{\otimes N} \, \frac{w(z_1)}{z_1}\] \cdot \[ (1+\hD)^{\otimes N}
 \,z_2\,w(z_2) \]     \,. \la{idgensup}
\end{equation}
For supergroups the diagramatics used in section \ref{diagramatics}
is still of great help but to avoid confusing various minus signs we
shall never use any component expression like \eq{1spins} or
\eq{2spins}. To read $\hD^N w(z)$ and $(1+\hD)^N w(z)$ we draw all
possible $N!$ permutation diagrams where the left-going lines from
upper indices to the lower indices are drawn as solid lines whereas
the right-going lines are dashes -- see figures
\ref{fig:BR1},\ref{fig:BR3} and \ref{fig:BR4}; for $\hD^N$ the
vertical lines are solid while for $(1+\hD)^N$ they are dashed.
Then, for each diagram, we first construct the tensor product
\begin{equation}
\frac{(gz)^{\epsilon_1}}{1-gz} \otimes \frac{(gz)^{\epsilon_2}}{1-gz}
\otimes \dots \otimes \frac{(gz)^{\epsilon_N}}{1-gz}
\end{equation}
where $\epsilon_n=1(0)$ if the line ending at the lower index $j_n$ is solid
(dashed). Finally we multiply this object by a product of superpermutation
operators which we read from the diagram\footnote{Notice that this procedure
of associating a set of super permutations to a given graph is completely
well defined since the super permutation obeys the same set of relations as
the usual permutator -- see for example \eq{13etc}. Consider for example the
third term in the \eq{fig3eq}. It corresponds to the third diagram in figure
\ref{fig:BR3}. We could associate to this diagram the permutation
$\mathcal{P}_{13}$ (by considering the vertical line as spectator) or
$\mathcal{P}_{12} \mathcal{P}_{23} \mathcal{P}_{12}$ (by slightly pushing
the vertical line to the left and accounting for the three interceptions
with this line) or $\mathcal{P}_{23} \mathcal{P}_{12} \mathcal{P}_{23}$ (by
shifting the vertical line slightly to the right and accounting for the
three interceptions with this line). These three possibilities are indeed
the same due to \eq{13etc} which is nothing but the YB relation for the
fundamental $R$-matrices at zero spectral parameters. }. As an example let
us write explicitly the 6 diagrams of figure \ref{fig:BR3}:
\begin{eqnarray}
\hD^N w(z)&=&\(\frac{gz}{1-gz} \otimes \frac{gz}{1-gz} \otimes \frac{gz}{1-gz}\) +\mathcal{P}_{12} \(\frac{1}{1-gz} \otimes \frac{gz}{1-gz} \otimes \frac{gz}{1-gz} \)\\
&+&\mathcal{P}_{13}\(\frac{1}{1-gz} \otimes \frac{gz}{1-gz} \otimes \frac{gz}{1-gz}\) +\mathcal{P}_{23} \(\frac{gz}{1-gz} \otimes \frac{1}{1-gz} \otimes \frac{gz}{1-gz} \)\\
&+&\mathcal{P}_{13}\mathcal{P}_{23}\(\frac{1}{1-gz} \otimes \frac{1}{1-gz}
\otimes \frac{gz}{1-gz}\) +\mathcal{P}_{23} \mathcal{P}_{12}
\(\frac{1}{1-gz} \otimes \frac{gz}{1-gz} \otimes \frac{gz}{1-gz} \)
\la{fig3eq}
\end{eqnarray}

Then, following the same reasoning as for the bosonic case we can prove the
identity \eq{idgensup} by establishing\footnote{We use a bosonic element $g$
(it is a group element) in the sense that $g=g^{ij} e_{ij}$ with $g^{ij}$
with fermionic (bosonic) grading for fermionic (bosonic) generators
$e_{ij}$. Then we can super permute $gz$ trivially -- see discussion bellow
\eq{bosonic} about the super permutation of states with even total grading.
That is, $\cP gz \otimes 1=1\otimes gz \cP$ etc. Thus formulae
\eq{id1sup},\eq{id2sup} can be trivially checked graphically -- see figure
\ref{fig:BR5} where the three spin example makes the general case obvious.}
\begin{eqnarray}
\hD^{N} w(z)&=&\[\(g \otimes \one \otimes \dots \otimes \one\) U_L \]
\,(1+\hD)^{\otimes N} z w(z) \,, \la{id1sup} \\
(1+\hD)^{N} w(z')&=&\hD^{\otimes N} \,\frac{w(z')}{z'}
\[\(g \otimes \one \otimes \dots \otimes \one\) U_L \]^{-1}\,. \la{id2sup}
\end{eqnarray}
where $U_L\equiv \mathcal{P}_{12}\mathcal{P}_{23}\dots \mathcal{P}_{L-1,L}$
is now the super shift operator.

Thus our derivations can be trivially generalized to include the supergroup
case as explained in this section and thus allow one to prove the BR formula
for the superalgebras $gl(K|M)$ with the twist element $g$.

\section{R-matrices in arbitrary irreps and commutativity of T-matrices} \la{Rcomsec}

Here we show how to construct the $R$-matrix in arbitrary symmetric irrep
$\lambda=1^s$
\begin{equation}\label{RMATR2}
    R_s(u)=u+2\sum_{\a\b}(-)^{p_\a}e_{\b\a}  \otimes  \pi_s(e_{\a\b})
\end{equation}
knowing the elementary $R$-matrix
\begin{equation}\label{ELEMR2}
    R(u)=u+2\cP_s=u+2\sum_{\a\b}(-)^{p_\a}e_{\b\a}  \otimes  e_{\a\b}
\end{equation}
and to prove the commutativity
\begin{equation}
[T_s(u),T_{s'}(u']=0 \la{TTss}
\end{equation}
of $T$-matrices in these  irreps. Obviously, in virtue of BR formula
\eq{BRF}, once this is proven we immediately get
\begin{equation}
[T_{\{\lambda\}}(u),T_{\{\lambda'\}}(u')]=0
\end{equation}
for any two representations\footnote{We should stress than in the process of
the derivation of the BR formula we used the fact that the transfer matrices
in the symmetric representation commuted but we never needed to use the
commutation of $T_{\{\lambda\}}$ for a generic representation.}.
\begin{figure} [t]   \centering
        \resizebox{151mm}{!}{\includegraphics{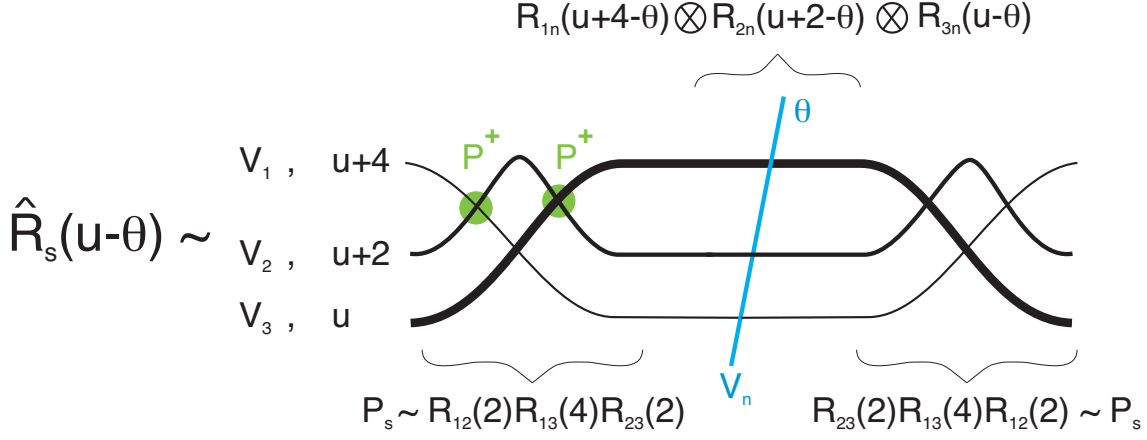}}
   \caption{ \footnotesize
   To build the $\hat R$-matrices in any symmetric representation from
   the fundamental $R$-matrices we start by drawing $s$ horizontal lines
   to which we associate the spectral parameters $u, u+2, \dots u+2s-2$
   and the vector spaces $V_s,V_{s-1},\dots, V_1$ -- all isomorphic to
   the fundamental vector space $V^{(K|M)}$ -- in vertical order. In the
    figure we represent the $s=3$ case.  To keep track of which line is
    which we represent the space $V_1$ by a thin line, $V_2$ by a thicker
    line, $V_3$ by an even thicker line etc. Then we cross these lines in
     all possible $\frac{s(s-1)}{2}$ ways. Each cross corresponds to an
      $R$-matrix acting on $V_i \otimes V_j$ with spectral parameter
      $u_j-u_i$ where $u_i,V_i$ and $V_j,u_j$, with $i>j$, are associated
       which each of the lines. In this way we build the first projector $P_s$.
The three lines are then crossed by a fourth line corresponding to
the physical space $V_n$ on which $\hat R_{s}$ also acts. The $s$
intersections with this new line correspond to a factor of $s$
fundamental $s$ matrices $R_{s,n}(u-\theta)\otimes \dots \otimes
R_{1,n}(u+2s-2-\theta)$. Finally the $s$ out-going lines are crossed
again in all possible ways and this gives us the remaining projector
$P_s$ in \eq{CONSTR}. Notice that the order by which the lines are
crossed is irrelevant due to the YB relation for the fundamental
$R$-matrices.}\label{fig:TTR}
\end{figure}

According to the general recipe \cite{KulResh1,Cherednik} we construct
$R_s(u)$ as follows:
\begin{equation}
R_s(u) S(u)=\hat R_s(u) \equiv  P_s\[R(u)\otimes R(u+2)\otimes \dots\otimes
R(u+2s-2)\]P_s \la{CONSTR}
\end{equation}
where $S(u)$ is a polynomial with fixed zeroes, to be precised bellow. In
what follows we shall check that this procedure does lead to \eq{RMATR2}.
The $R$-matrices inside the brackets are multiplied only in the quantum
space. Since the $R$-matrices degenerate at some special points
\begin{equation}
    R(\pm 2) \sim \cP_\pm
\end{equation}
into symmetric and anti-symmetric projectors, the symmetric projectors $P_s$
can be constructed out of products of $R$-matrices at special points. The
rule to construct $P_s$  is the following: one crosses $s$ lines of the
object \eq{CONSTR} in all possible $\frac{s(s-1)}{2}$ ways, associating the
corresponding $R$-matrices to the crossings as explained in figure
\ref{fig:TTR} where the $s=3$ case is depicted.

\begin{figure} [h]   \centering
        \resizebox{151mm}{!}{\includegraphics{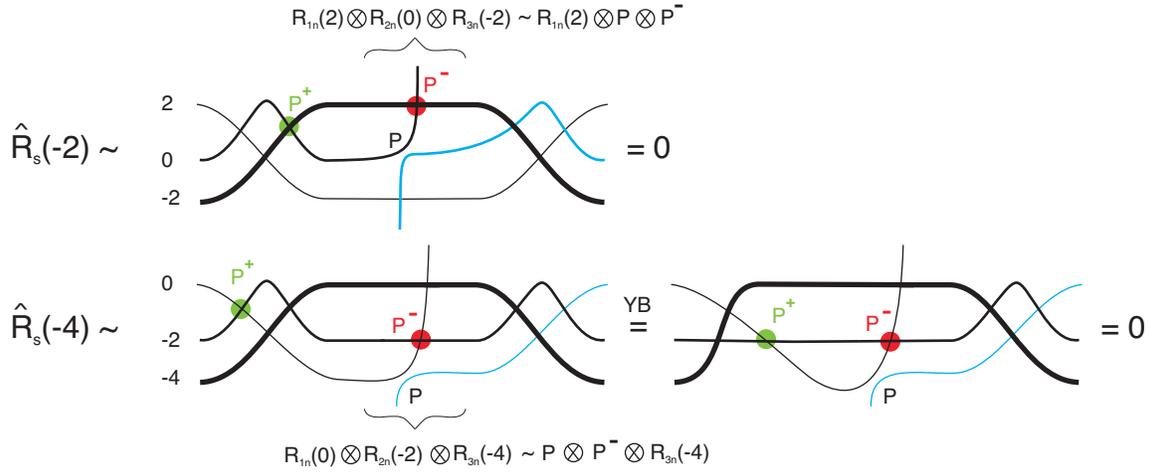}}
\caption{\footnotesize The hatted $R$-matrix acts on $V^{1^s}\otimes
V^{(n)}$ where $V^{(n)}$ is some quantum space in the fundamental
representation. As described in figure $6$ it is given by $
\hat{R}_s(u)=\prod\limits_{j,i<j}^{s} R_{ij}(2(j-i))
\prod\limits_{j=1,\dots,s} R_{jn}(u+2(s-j)) \prod\limits_{j,i<j}^{s}
R_{ij}(2(j-i)) $. Thus at $u=2,\dots,2s-2$ it will contain a factor
$R_{j,j+1}(2) R_{j+1,n}(0)R_{j,n}(-2)\sim
\mathcal{P}^+_{j,j+1}\mathcal{P}^{}_{j+1,n}\mathcal{P}^-_{j,n}=
\mathcal{P}^+_{j,j+1}\mathcal{P}^-_{j,j+1} \mathcal{P}^{}_{j+1,n}=0$
so that the $R$-matrix is zero at these points. In the figure this
phenomena is demonstrated for $s=3$.}\label{fig:Zeros}
\end{figure}

When $u=-2,-4\dots,-2s+2$ we will find the combination
\begin{equation}
\mathcal{P}^+_{12}\mathcal{P}^{}_{23}\mathcal{P}^-_{13}=0
\end{equation}
in $\hat {R}_s(u)$ which will therefore vanish at these points -- see figure
\ref{fig:Zeros}. Thus $\hat{R}_s(u)/S(u)$ with $S(u)= \prod_{k=1}^{s-1}
(u+2k) $ must be a linear polynomial in $u$.

Hence, to establish \eq{CONSTR}, we merely have to check that this
linear polynomial coincides with $R_s(u)$ in \eq{RMATR2}. The linear
term is obviously equal to $u \one\otimes P_s$, where $\one$
corresponds to the (super)vector quantum spin space and $P_s$
corresponds to the auxiliary space, whereas the  other term can be
extracted from the $u\to\infty$ limit of r.h.s. of \eq{CONSTR}. The
result is easily seen to be
\begin{equation*}
    R_s(u) = u \one\otimes P_s+ 2 \sum_{\a\b}(-)^{p_\b}e_{\b\a}
     \otimes  P_s\[ e_{\a\b}\otimes\one\dots\otimes\one
  + \one\otimes e_{\a\b}\otimes\one\dots\otimes\one +\cdots
  + \one\dots\otimes\one\otimes e_{\a\b} \]P_s
\end{equation*}
The expression in square brackets containing $s$ terms, surrounded by two
symmetric $P_s$-projectors, is precisely the generator $\pi_s\(e_{\a\b}\)$
in symmetric $1^s$ irrep. One can check for example that it satisfies the
usual $gl(K|M)$ commutation relations for the generators.
 Hence we proved that
\eq{RMATR2} is indeed the $R$-matrix mixing the quantum vector
representation with the auxiliary symmetric irrep. Thus the procedure
\eq{CONSTR} does allow one to fuse the fundamental $R$-matrices \eq{ELEMR2}
into the $R$-matrices in an arbitrary symmetric irrep $\l=1^s$.

\begin{figure} [h]   \centering
        \resizebox{151mm}{!}{\includegraphics{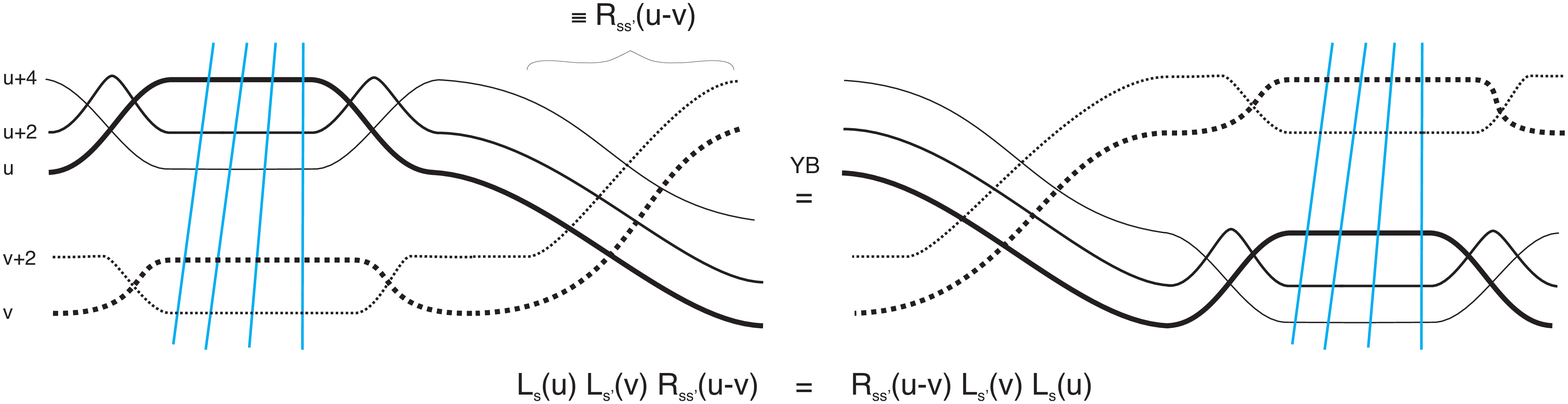}}
   \caption{ \footnotesize If we cross all outgoing lines from $L_s(u)$
   and $L_{s'}(v)$ we obtain some object acting on $V^{1^s}\otimes V^{1^{s'}}$
   which we denote by $R_{ss'}(u-v)$ -- see figure where $s=3$ and $s'=2$.
   But then, due to the YB relations for the fundamental $R$-matrices,
   that is the crosses in the figure, we can shift any lines through any
   other lines and in particular we see that $L_{s}(u)L_{s'}(v)R_{ss'}(u-v)
   =R_{ss'}(u-v)L_{s'}(v)L_{s}(u)$. Multiplying this relation by
   $R_{ss'}^{-1}(u-v)$ and taking the trace we obtain the desired
   relation $\[T_{s}(u),T_{s'}(v)\]=0$ for the transfer matrices in the
   symmetric representations.}\label{fig:RTT}
\end{figure}

Next, to check \eq{TTss} it suffices to notice that
\begin{equation}\label{TTR}
    L_{s'}(u')L_{s}(u)R_{s',s}(u-u') =R_{s',s}(u-u')L_s(u)L_{s'}(u')
\end{equation}
where $R$-matrix $R_{s',s}(u'-u)$  intertwinning two symmetric irreps is
\textit{by definition} the appropriate product of all $R$-matrices arising
in the intersections of the  lines  of two auxiliary spaces $s$ and $s'$ and
$L_{s}$ is the monodromy matrix for the $1^s$ representation (see fig
\ref{fig:RTT}).
Multiplying  this expression by $R^{-1}_{s',s}(u-u')$ and taking the
trace we find \eq{TTss}  as announced in the beginning of this
section.

\section{Hirota relation }

The BR formula \eq{BRF} allows to take an interesting approach to the
quantum integrability, including the diagonalization of transfer-matrices
and related hamiltonians of quantum spin chains, and  the retrieval of
Baxter equations and Bethe ansatz equations by the use of integrable
classical discrete dynamics \cite{KLWZ,KSZ,Zabrodin}. We will briefly
describe here the Hirota equation and the curious identities for the
characters induced by it.

Hirota identity concerns the specific  irreps $\l=a^s$ with the
rectangular Young tableaux (YT) of the size $a\times s$. The
(super)characters, due to the  determinant representation
\eq{CHADET}, obey the following Hirota identity:
\begin{equation}\label{CHAHI}
    \chi^2(a,s)=\chi(a+1,s)\chi(a-1,s)+\chi(a,s+1)\chi(a,s-1)
 \end{equation}
It shows that the (super)characters represent the $\tau$-functions
of the discrete KdV-hierarchy \cite{JimboMiwa}.

The super-characters are nonzero only for the YT's in the "fat hook" region,
where all the highest weights are positive and $\la_{K+1}\le M$. The
following identity can be proven
  \begin{equation*}
     \chi(K,M+n)=\prod_{k=1}^K x_k^n\, \prod_{i=1}^K\prod_{j=1}^M (x_i-y_j)
     =(-)^{nM}({\rm sdet~})^n\chi(K+n,M)
  \end{equation*}
  which shows that these representations, lying on horizontal and vertical interior boundaries of the fat
  hook, are identical. They are called typical, or long
  representations. They can be analytically continued to the non-integer values of n. The rest of the
rectangular representations are called atypical, or short.

 Applying the Jacobi identity for determinants -- see appendix A.3 -- we can
 easily show
 that \eq{BRF} in case of the rectangular irrep $\l=a^s$ implies the Hirota
 equation for the $T$-matrix\footnote{See the mathematical papers \cite{Hernandez1,Hernandez2} for its mathematical demonstration}
 \begin{equation}\label{THI}
    T(u+1,a,s)T(u-1,a,s)=T(u+1,a+1,s)T(u-1,a-1,s)+T(u-1,a,s+1)T(u+1,a,s-1)
  \end{equation}
  Let us plug into this equation our representation \eq{DIFFT} for the
  irrep $\l=a^s$, first  in case of
  the   one spin chain. We obtain
  \begin{eqnarray*}
   \(u+1 +2\hD\)\chi(a,s)\(u-1 +2\hD\)\chi(a,s) &=&\nn\\
   \(u+1 +2\hD\)\chi(a+1,s)\(u-1 +2\hD\)\chi(a-1,s)
   &+&\(u-1 +2\hD\)\chi(a,s+1)\(u+1 +2\hD\)\chi(a,s-1)
 \end{eqnarray*}
 Now note that the terms proportional to $u^2-1$, containing no derivative $\hD$,
 cancel due to \eq{CHAHI}. The terms proportional to $u$ contain only one $\hD$
 each.
  They  combine into the derivative of Hirota relation  $\hD$[\eq{CHAHI}] and thus
  cancel as well. {\it The $u$-dependent terms cancel!} Taking  $u=1$, we are left
  only with the $u$-independent identity on characters to check:
\begin{eqnarray}\label{FINALID}
 \nn  \(1 +\hD\)\chi(a,s)\cdot \hD \chi(a,s) =
 \(1 +\hD\)  \chi(a,s+1) \cdot
   \hD \chi(a,s-1)+
   \nn
    \(1 +\hD\)  \chi(a-1,s) \cdot
   \hD  \chi(a+1,s)
 \end{eqnarray}
Here in each term $\hD$ acts only on the following character. It would be
nice to relate this identity to the Hirota equation for the discrete
KdV-hierarchy. The role of the evolution "times" should be played then by
$t_q=\tr g^q$. For finite rank $K$ only $K$ first "times" are independent.

In the case of arbitrary number $N$ of spins Hirota relation \eq{THI} takes
the form:
\begin{eqnarray*}
   \hat L_N^+\chi(a,s)
   \hat L_N^-\chi(a,s)
   -\hat L_N^+\chi(a,s+1)
   \hat L_N^-\chi(a,s-1)
   -\hat L_N^+
   \chi(a-1,s)\hat L_N^-
   \chi(a+1,s)=0
 \end{eqnarray*}
where we introduced the operators
\begin{eqnarray*}
   \hat L_N^\pm = \(u_1\pm 1 +2\hD \)\otimes\dots\otimes \(u_N\pm 1
   +2\hD \).
 \end{eqnarray*}
 We obtain a set of relations on characters. Since they are true for any
 values of $u_1,\cdots,u_N$ we can choose any particular values.
 The choice $u_1=u_2=\dots=u_N=1$ for the chain of $N$ spins gives the
 following curious
  identity\footnote{Indeed it is trivial to check that for the identity with
   $N$ spins all $u$ dependent terms are trivial by virtue of the identity for
   $N-1$, following the same recurrence reasonings involved in section
    \ref{secgen}.}:
\begin{eqnarray} 
   &&(1 +\hD)^{\otimes N}\chi(a,s)\cdot \hD^{\otimes N}\chi(a,s) =
   \nn\\
   &&(1 +\hD)^{\otimes N} \chi(a,s+1) \cdot
   \hD^{\otimes N} \chi(a,s-1)+
   \nn
    (1 +\hD)^{\otimes N} \chi(a-1,s) \cdot
   \hD^{\otimes N} \chi(a+1,s)
 \end{eqnarray}
Probably, this identity on characters,  generalizing the one for symmetric
characters \eq{SPINIDgen}, can be also viewed as a special case of Hirota
equation for the tau function (character) of the discrete KdV Hierarchy.
This opens a paradoxical possibility to interpret the quantum integrable
systems as a particular case of classical integrable systems with discrete
dynamics. In the integrable world quantization rather means discretization.

\section{Discussion}

In this paper, we derived in a rather straightforward way the
Bazhanov-Reshetikhin relations for transfer matrices of integrable
(super)spin chains. Our starting point is the most basic object - the
rational $R$ matrix on $gl(K|M)$ superalgebra. The corresponding transfer
matrix is twisted by a general $GL(K|M)$ element. Our method is closely
related to the usual $GL(K|M)$ characters. It is natural since the
BR-formula is the direct  generalization of the Jacobi-Trudi determinant
formula for the (super)characters to the case of quantum characters -
transfer matrices.

The   $GL(K|M)$ twist is an important ingredient in our construction. It
allows to avoid the use of such complicated objects as  superprojectors. We
work directly with the transfer-matrix, and not with the monodromy matrix,
so the indices of auxiliary space are always contracted. The indices of the
quantum space are opened, spin by spin, by means of  the convenient group
derivatives acting on usual (super)characters, to produce the quantum
characters - transfer-matrices. The action of these derivatives on
generating functions of characters is relatively simple. In addition,  this
twist is a natural regularization of quantum transfer matrices, instead of
the less invariant Cherednik's regularization of projectors to irreps, built
out of elementary $R$-matrices \cite{Cherednik}.

Our method might be useful to advance  the understanding of quantum
integrability. For example, one could try, using our formalism, to derive
the B\"acklund relations of the paper \cite{KSZ}, presented the
in section 8, directly from our representation \eq{DIFFT}, using the
Gelfand-Zeitlin reduction. In that case, we would restore the operatorial
meaning of Baxter functions, at every step of nesting of the type
$gl(K|M)\supset gl(K-1|M)\supset gl(K-1|M-1)\dots gl(1|1)\supset
gl(1|0)\supset \emptyset$. This would be  probably the  most direct shortcut
to the nested  Bethe ansatz equations diagonalizing the transfer matrices.

The method can be helpful to attack more complicated systems. In
particular, the  generalization of our derivation of the
Bazhanov-Reshetikhin formula to the  trigonometric $R$-matrices
(quantum groups) and to the elliptic $R$-matrices\footnote{The
latter are not known in the supersymmetric case} would be
interesting to establish. The analogues of BR formula for the
$so(N)$, $sp(N)$ and $osp(m|2n)$ algebras would be also interesting
to derive by our method. For these Lie algebras the conjectured BR
formula is not a simple spectral parameter dependent version of the
Jacobi-Trudi type formulas. Thus, in this case, the generalization
of our quantization procedure would probably require to consider the
action of our co-derivative on (linear) combinations of the
characters of the classical algebra. Another important direction
would be the inclusion of non-compact representations of
(super)groups in our approach. The formalism of
\cite{Belitsky,Belitsky2} could be a good starting point for it.

On other algebras (so(N), sp(N),,,,), one can not get
Bazhanov-Reshetikhin-like formula just by putting spectral parameter into
 the Jacob-Trudi type formula on classical algebras.
In this sense, the fact that one can get Bazhanov-Reshetikhin formula
for gl(m|n) just by putting spectral parameter into the (classical) Jacob-Trudi
formula is an almost accident.
To extend your method to other algebras, you will have to consider
"group derivatives" on linear combinations of characters of classical
algebras.

It would be also very interesting to understand the connection between our
formalism and the Cherednik/Drinfeld duality in the lines of \cite{CD1,CD2,CD3}.
 In particular the relation between our co-derivative and the Cherednik/Drinfield
  composite functor described e.g. in \cite{CD3} would be worth exploring\footnote{We thank M. Nazarov for calling our attention to this interesting connection.}.

Even more interesting would be to treat by this method the
transfer-matrices based on "non-trivial" R-matrices, like the one
for Hubbard chain \cite{Shastry} and the recently constructed
AdS/CFT $R$-matrix with the symmetry of $sl(2|2)$ supergroup
extended by central charges \cite{Mathias,Beisert1,Beisert2,Arutyunov:2006yd}.

Another interesting question concerns various classical limits of the
quantum (super)spin chains in this language. This limit usually corresponds
to  large values of the spectral parameter (low lying energy levels of the
system), the  large number of spins and of magnon excitations. It
 is known to be very direct and transparent   from
the Baxter-type TQ-relations between transfer matrices and Baxter's $Q$
functions (see for example \cite{BKSZII}). The complete set of such relations
for $gl(K|M)$ rational case, as well as the new $QQ$ type relations, is
available from \cite{KSZ}. But it would be interesting to extract the
classical limits directly from the Bazhanov-Reshetikhin formula \eq{BRF}.

A very interesting route to explore, using our approach, is the connection
of integrability of quantum (super)spin chains to the classical integrable
hierarchies. It stems from the striking observation that the {\it
quantization} in the integrable world often means {\it discretization}.
Indeed, in the approaches of \cite{KLWZ,KSZ} the quantized spin chain was
represented by the integrable Hirota equation for its quantum
transfer-matrix eigenvalues. Our  approach based on characters and their
quantum generalization sheds more light on this unusual "classical"  nature
of quantum integrable models (very different from various classical limits
of the same models). Already the simple (super)character represents the
tau-function of the discrete KdV hierarchy. The identities for characters
obtained from the full quantum Hirota relation for fusions at the end of the
section 7, are probably a particular form of Hirota relations for the
discrete KdV tau-functions. The  evolution "times" are related to the values
of the twist matrix $g$.

One more interesting "classical" limit to study could here be the
large rank of the (super)group: $K,M\to\infty$. It should probably
be accompanied by the limit of big irreps, or  big young tableaux in
the auxiliary space. This is the closest analogue of the large $N$
limit in matrix models, since the character itself can be viewed as
a unitary one-matrix integral \eq{INTCHAR}. A good starting point
here is the large $N$ limit for characters investigated in
\cite{Kazakov:1995ae,Kazakov:1996zm}. Many random matrix techniques
could be applicable here, and this link to the quantum integrability
can significantly and profoundly enrich the subject of random
matrices itself.

\section*{Acknowledgements}

We would like to thank N.~Beisert, I.~Cherednik, N.~Gromov, I.~Kostov,
P.~Kulish, J.~Minahan, M.~Nazarov, J.~Penedones, P.~Ribeiro, D.~Serban,
A.~Sorin, V.~Tolstoy, Z.~Tsuboi, P.~Wiegmann, A.~Zabrodin and K.~Zarembo for
discussions at different stages of this work. The work of V.K. has been
partially supported by European Union under the RTN contracts
MRTN-CT-2004-512194 and by the ANR program INT-AdS/CFT -ANR36ADSCSTZ. P.~V.
is funded by the Funda\c{c}\~ao para a Ci\^encia e Tecnologia fellowship
{SFRH/BD/17959/2004/0WA9}. V.K. thanks the Banff Center for Science
(Canada), Physics department of Porto University (Portugal) and the Max
Planck Institute (Potsdam, Germany), where a part of the work was done, for
the hospitality. The visit of V.K. to Max Planck Institute was covered by
the Humboldt Research Award.


\section*{Appendix A: (Super-)Characters}

We present here some, not exhaustive, but a self-consistent set of formulas
demonstrating  the Jacobi-Trudi (second Weyl formula) for $GL(K)$
characters, and then generalize them for the $GL(K|M)$ super-characters.

\subsection*{A.1 Definition, generating function and integral representation}

A general  element $g\in GL(K,{\cal R})$ can be represented as
\begin{equation}\label{GELEM}
    g=\exp\[\sum_{\a,\b=1}^K  e_{\a\b}\phi_{\a\b}\]
\end{equation}
where $\phi_{\a\b}$ is a $K\times K$ matrix of real numbers and the $K^2$
generators $e_{\a\b}$ satisfy the commutation relations
\begin{equation}\label{GENER}
    \[e_{\a_1\b_1},e_{\a_2\b_2}\]=\delta_{\b_1\a_2}e_{\a_1\b_2}
    -\delta_{\a_1\b_2}e_{\a_2\b_1}
\end{equation}
In the simplest case of fundamental representation
$\[\,e_{\a\b}\]^{i}_{j}=\delta^i_\a\delta_{j,\b}$ and $g=e^\phi$.

For a more general representation $\l$ the generators $e_{\a\b}$ take values
in a larger  vector space characterizing the representation.
 The irreducible representations ("irreps") $\l$ of $GL(K,{\cal R})$
(or rather of its positive signature component $GL_+(K,{\cal R})$) are
characterized by the highest weight components: the ordered non-negative
integers: $\l=\(\l_1\ge\l_2,\dots,\ge\l_{K}\)$. They are isomorphic to the
corresponding unitary irreducible representations of the group $U(K)$
(limited to the positive highest weight components). Hence we can construct
the matrix elements $\pi_\l(g)$ and the  characters
$\chi_\l(g)=\tr\pi_\l(g)$ of a group element $g$ for $U(K)$ and then
analytically continue them to $GL(K,{\cal R})$.

Now, given two representations $\lambda$ and $\lambda'$ the group element
$g$ in these representation obey the standard orthogonality condition
\begin{equation}
\int dg \,\pi_\lambda(g) \otimes \pi_{\lambda'}(g^{-1})=\frac{1}{d_\l}
\delta_{\l\l'} \mathcal{P}_\l \la{ORTHO}
\end{equation}
where $dg$ is the invariant Haar measure on the group $U(K)$ normalized to
$1$, $\mathcal{P}_\lambda$ is the permutation operator acting on
$V_{\lambda}\otimes V_{\lambda}$ and $d_\l$ is the dimension of the
representation $\l$. The completeness condition reads
\begin{equation}
\sum_{\l} d_\l\,{\rm tr}_\l \[\,\pi_\lambda\(g\) \pi_{\lambda}({g'}^{-1} )\]
=\delta\(g- g'\)
\end{equation}
If we multiply \eq{ORTHO} by $\pi_\l(h)\otimes 1$ and trace over the second
 space we get
\begin{equation}
\int dg \,\pi_\lambda(h g) \chi_{\lambda'}(g^{-1})=\frac{1}{d_\l}
\delta_{\l\l'}
 \pi_\l(h)
\end{equation}
and if we take the trace of this expression we obtain
\begin{equation}
\int dg \,\chi_\lambda(h g) \chi_{\lambda'}(g^{-1})=\frac{1}{d_\l}
\delta_{\l\l'} \chi_\l(h)
\end{equation}
which   reduces for $h=1$ to the simple character orthogonality condition
\begin{equation}
\int dg \,\chi_\lambda(g) \chi_{\lambda'}(g^{-1})= \delta_{\l\l'} \,.
\la{ORTHOchi}
\end{equation}
Indeed, any invariant function on the $U(K)$ group $f(g)=f(\O^\dagger
g\O),\,\,\, \O\in U(N)$, can be expanded into the "Fourier" series w.r.t.
the characters over all irreps
\begin{equation*}
    f(g)=\sum_{\l}C_\l\chi_\l(g)
\end{equation*}
with
\begin{equation*}
    C_\l=\int dg f(g)\chi_\l(g^\dagger) \,.
\end{equation*}

Also, the  following completeness property takes place
\begin{equation}\label{COMPLET}
   \det \(1- h \otimes g\)^{-1}=\sum_{\l}\chi_\l(g)\chi_\l(h)
\end{equation}
which can easilly be checked by multiplying both sides of this identity by
any group invariant function $f(h^{-1})=\sum_{\l}C_{\l} \chi_{\l}(h^{-1}) $
and integrating over $h$ with the Haar measure. From the r.h.s we get
\begin{equation}
\int dh \sum_{\l\l'}\chi_\l(g)\chi_\l(h) \, C_{\l'} \chi_{\l'}(h^{-1})
=f(g)
\end{equation}
where the orthogonal relation \eq{ORTHOchi} was used. From the l.h.s.
\begin{equation}\label{CHICHI}
\int dh \det \(1- h \otimes g\)^{-1} f(h^{-1})=f(g)
\end{equation}
with the integral  calculated by "poles" $h=g^{-1}$.

We can clarify it if we go to the eigenvalues: $g=\O^\dagger X\O$,
$X=diag\{x_1,x_2,\dots,x_K\}$, and similarly for $h=\tilde\O^\dagger
Z\tilde\O$, $Z=diag\{z_1,z_2,\dots,z_K\}$, when the completeness condition
becomes
\begin{equation}\label{XCOMPLET}
   \prod\limits_{a,b=1}^K \(1- x_a z_b\)^{-1}=\sum_{\l}\chi_\l(X)\chi_\l(Z)
\end{equation}

Let us show that this condition, accompanied by the corresponding
analyticity properties is satisfied by the characters given in terms of the
2-nd Weyl formula:
\begin{equation}\label{WEYL}
   \chi_\l(X)=\frac{\det\limits_{1\le i,j\le K}x_i^{\l_j-j }}{\Delta(x_1,\dots,x_K)}
\end{equation}
where $\Delta(x_1,\dots,x_K)=\prod\limits_{a<b}\(x_a- x_b\)$. Indeed
plugging this into \eq{XCOMPLET} we obtain precisely the Cauchy identity.

Now, using \eq{CHICHI}, we write the integral representation for the
character:
\begin{equation*}
   \chi_\l(X)=\oint \prod\limits_{k=1}^K dz_k \, \Delta\(z_1,\dots,z_K\)
   \frac{\det\limits_{1\le i,j\le K}x_i^{-\l_j+j }}{\prod\limits_{a,b=1}^K \(1- x_a z_b\)}
\end{equation*}
The integration contours here go around the origin, avoiding the
singularities of the denominator. If all $\l_k=0,\,\,\,  a< k\le K$ one can
show that the last formula  becomes
\begin{equation} 
    \chi_\l= \frac{1}{a!}\oint \prod_{1\le n\le a}
    \frac{dt_n\, w(t_n)}{2\pi i\, t_n^{1+\l_n}}   \Delta(t_1,\dots,t_a)
\end{equation}
where the contours go around the concentric unit circles, and
\begin{equation*}
    w(t)=\det \(1-t g\)^{-1}
    =\frac{1}{\prod\limits_{k=1}^K(1-x_k t)}
    =\sum_{s=1}^\infty \chi_s t^s= \( \sum_{a=1}^\infty \chi^a t^a  \)^{-1}
\end{equation*}
is the generating function of characters of symmetric irreps (Schur
functions) $\chi_s$ and of antisymmetric irreps $\chi^a$. For the specific
irreps $\l=a^s$ with the rectangular Young tableaux  of the size $a\times
s$.
\begin{equation*}
    \chi(a,s)= \int \frac{[d\,h]_{\rm SU(a)}}{ \(\det h\)^{s+1}}\,\,\,
    \det\(1-h\otimes g\)^{-1}=\frac{1}{a!}\oint \prod_{1\le n\le a}
    \frac{dt_n\, w(t_n)}{2\pi i\, t_n^{1+s}} \left|\Delta\(t_1,\dots,t_{a}\)
    \right|^2
\end{equation*}
Expanding and picking up the poles of the denominator $t_a=\frac{1}{x_b}$ we
arrive at the Jacobi-Trudi formula for characters.
\begin{eqnarray}\label{JACOBITRUDI}
 \chi_{\{\l\}}(g) = \det_{1\le i,j\le a}\chi_{\l_j+i-j} (g)\,.
\end{eqnarray}
For the characters of rectangular irreps $\l=a^s$ ,  the following formula
follows from it
\begin{eqnarray*}
 \chi(a,s) = \det_{1\le i,j\le a}\chi(1,s+i-j)= \det_{1\le i,j\le s}\chi(a+i-j,1)
\end{eqnarray*}

\subsection*{A.2 Generalization to super-characters   }
\begin{figure} [t]   \centering
        \resizebox{100mm}{!}{\includegraphics{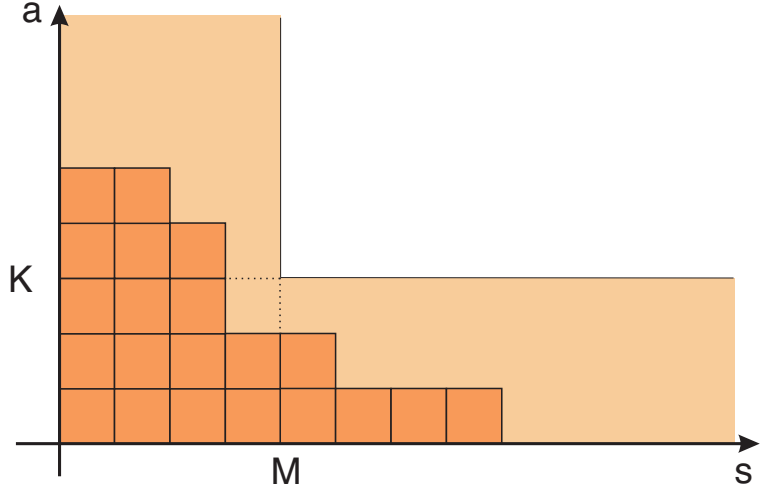}}
   \caption{ \footnotesize
For the supergroups $GL(K|M)$ the young tableaux can be as infinitely big
provided they stay inside the fat hook region as indicated in the
figure.}\label{fig:chi}
\end{figure}
The irreps $\l$ of the supergroup $GL(K|M)$ are described by similar Young
tableaux as the irreps of the usual $GL(K)$, and are characterized by the
highest weight components $\l=\{\infty
>\l_1\le\l_2\le\dots\le \l_n\dots\}$ but number of these components is not
restricted. The only restriction on the shape of these super-tableaux is  on
the $K+1$-th highest weight: $\l_{K+1}\le M$. This limits the allowed Young
super-tableaux to the "fat hook" domain presented on figure \ref{fig:chi}.

For super-groups the Jacobi trudi formula \eq{JACOBITRUDI} remains valid but
there are much larger family of representations (typical, atypical etc) and
there is no general Weyl formula as for the bosonic groups. The symmetric
functions are now given by the generating function
\begin{equation}\label{SSCHUR}
    \sw(t)={\rm s}\!\det \(1-t g\)^{-1}
    =\frac{\prod_{m=1}^M(1-y_m t)}{\prod_{k=1}^K(1-x_k t)}
    =\sum_{s=1}^\infty \schi(1,s)t^s= \( \sum_{a=1}^\infty \schi(a,1)t^a  \)^{-1}
\end{equation}
where we diagonalized the supermatrix $g \in GL(K|M)$ as
\begin{equation} 
\Omega g \Omega^\dagger={\rm diag}(x_1,\dots,x_K|y_1,\dots ,y_M)
\end{equation}
Then, as before, formula \eq{INTCHAR} holds,
\begin{equation}\label{INTCHAR}
    \chi_\l= \frac{1}{a!}\oint \prod_{1\le n\le a}
    \frac{dt_n\, w(t_n)}{2\pi i\, t_n^{1+\l_n}}   \Delta(t_1,\dots,t_a) \,,
\end{equation}
 and going here to the super-eigenvalues,  expanding as in \eq{SSCHUR}
and picking up the poles of the denominator $t_a=\frac{1}{x_b}$ we arrive at
the Jacobi-Trudi formula for super characters
\begin{eqnarray}\label{SUPERJACOBITRUDI}
 \schi_{\{\l\}}(g) = \det_{1\le i,j\le a}\schi_{\l_j+i-j} (g)\,.
\end{eqnarray}
This is not different from the one for usual $GL(K)$ characters and is still
given by \eq{JACOBITRUDI}, but the irreps are characterized by  a set of the
Young super-tableaux. The characters are nonzero only for the YT's in the
"fat hook" region, where all the highest weights are positive and
$\la_{K+1}\le M$ (see  \cite{Bars1,Bars2,Bars3} for the description).

\subsection*{A.3 B\"acklund relations for (super)characters}

\begin{figure} [t]   \centering
        \resizebox{151mm}{!}{\includegraphics{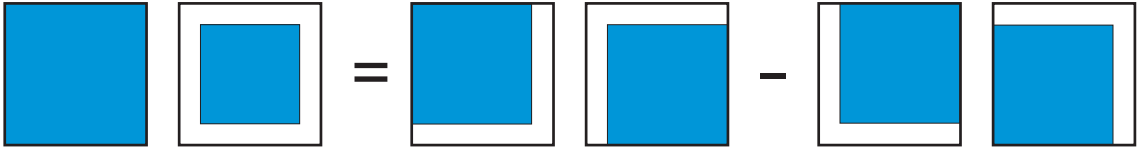}}
   \caption{ \footnotesize
The Jacobi identity \eq{JACOBI} for the determinants of a matrix $M$ and the
determinants of the same matrix with some lines or columns chopped out. The
full painted box represents the determinant of the matrix $M$ while the
second figure with a painted box inside the larger box means the determinant
of the matrix obtained from $M$ by taking out the first and last lines and
columns. The four pictures in the r.h.s. correspond to the determinant of
the matrix obtained from $M$ by removing the first/last line and the
first/last column in all possible $4=2\times 2$ combinations. For a
two-by-two matrix the Jacobi identity is nothing but the everyday formula
used to compute the determinant of the matrix $M$.}\label{fig:Jacobi}
\end{figure}

Let us now precise  the notations for the generating functions and the
characters of rectangular irreps on the supergroup $gl(K|M)$ as
$\chi_{K,M}(a,s)$ and  $w_{K,M}(t)$, respectively. From the definition
\eq{DEFW} we have obvious relations between the generating functions for the
groups of  different ranks
\begin{eqnarray}\label{WTW}
     w_{K-1,M}(t)&=&(1-t\,\, x_{K}) w_{K,M}(t)\nn\\
     w_{K,M}(t)&=&(1-t\,\, y_{M}) w_{K,M-1}(t)\nn
 \end{eqnarray}
 and hence, for the characters of symmetric irreps (Schur
 functions):
 \begin{eqnarray}\label{CHITCHI}
    \chi_{K-1,M}(1,s)&=&\chi_{K,M}(1,s)-x_{K}\chi_{K,M}(1,s-1)    \nn\\
     \chi_{K,M}(1,s)&=&\chi_{K,M-1}(1,s)-y_{M}\chi_{K,M-1}(1,s-1)\nn
 \end{eqnarray}
B\"acklund transformations for characters then follow \cite{Zabrodin}
\begin{eqnarray}\label{CHBACKLUND}
    \chi_{K,M}(a,s+1)\chi_{K-1,M}(a,s)-\chi_{K,M}(a,s)\chi_{K-1,M}(a,s+1)&=&
    x_K\,\chi_{K,M}(a+1,s)\chi_{K-1,M}(a-1,s+1)\nn\,,\\
        \chi_{K,M}(a+1,s)\chi_{K-1,M}(a,s)-\chi_{K,M}(a,s)\chi_{K-1,M}(a+1,s)&=&
    x_K\,\chi_{K,M}(a+1,s-1)\chi_{K-1,M}(a,s+1)\nn\,,\\
    \chi_{K,M-1}(a,s+1)\chi_{K,M}(a,s)-\chi_{K,M-1}(a,s)\chi_{K,M}(a,s+1)&=&
    y_M\,\chi_{K,M-1}(a+1,s)\chi_{K,M}(a-1,s+1)\nn\,, \\
        \chi_{K,M-1}(a+1,s)\chi_{K,M}(a,s)-\chi_{K,M-1}(a,s)\chi_{K,M}(a+1,s)&=&
    y_M \,\chi_{K,M-1}(a+1,s-1)\chi_{K,M}(a,s+1)\nn\,.
\end{eqnarray}
The proof of the first one e.g.: take the $(a+1)\times(a+1)$ matrix with
only the first column consisting of $\chi_{K,M}$'s, the rest - of
$\chi_{K-1,M}$'s
\begin{equation*}
    \left(
       \begin{array}{cccccc}\label{TTDET}
\chi_{K,M}(1,s)   & \chi_{K-1,M}(1,s) & \dots & \chi_{K-1,M}(1,s-j)
& \dots & \chi_{K-1,M}(1,s-a)  \\
\chi_{K,M}(1,s+1) & \chi_{K-1,M}(1,s+1)& \dots & \chi_{K-1,M}(1,s+1-j)
& \dots & \chi_{K-1,M}(1,s+1-a)  \\
\dots & \dots & \dots & \dots & \dots & \dots \\
\chi_{K,M}(1,s+a) & \chi_{K-1,M}(1,s+a) & \dots& \chi_{K-1,M}(1,s+a-j)
 & \dots & \chi_{K-1,M}(1,s+1)  \\
       \end{array}
     \right)
\end{equation*}
Applying the Jacobi identity (see figure \ref{fig:Jacobi})
\begin{equation}\label{JACOBI}
    D_{a+1}(m,n) D_{a-1}(m+1,n+1)=D_{a}(m,n) D_{a}(m+1,n+1)
    -D_{a}(m+1,n) D_{a}(m,n+1)
\end{equation}
for the determinants
$$
D_a(m,n)=\det\limits_{m+1\le i\le m+a,\,\, n+1\le j< n+a} M_{i,j}\,,
$$
where $M_{i,j}$ is any matrix, to the matrix written above, we obtain the
first Backlund transformation.


\end{document}